\pdfoutput=1
\documentclass[12pt,a4paper]{article}

\usepackage{ifthen} 
\newboolean{pdflatex}
\setboolean{pdflatex}{true} 

\newboolean{articletitles}
\setboolean{articletitles}{true} 

\newboolean{uprightparticles}
\setboolean{uprightparticles}{false} 

\def\paperauthors{LHCb collaboration} 
\def\paperasciititle{Model-independent measurement of charm mixing parameters} 
\def\papertitle{Model-independent measurement of charm mixing parameters in \SLDecay decays} 
\def\paperkeywords{{High Energy Physics}, {LHCb}} 
\def\papercopyright{\the\year\ CERN for the benefit of the LHCb collaboration} 
\def\paperlicence{CC BY 4.0 licence}
\def\paperlicenceurl{https://creativecommons.org/licenses/by/4.0/}

\usepackage[top=1in, bottom=1.25in, left=1in, right=1in]{geometry}

\usepackage{microtype}
\usepackage{lineno}  
\usepackage{xspace} 
\usepackage{caption} 

\usepackage{graphicx}  
\usepackage{color}
\usepackage{colortbl}
\graphicspath{{./figs/}} 

\usepackage{amsmath} 
\usepackage{amssymb}
\usepackage{amsfonts}
\usepackage{upgreek} 

\newcommand*\patchAmsMathEnvironmentForLineno[1]{%
\expandafter\let\csname old#1\expandafter\endcsname\csname #1\endcsname
\expandafter\let\csname oldend#1\expandafter\endcsname\csname
end#1\endcsname
 \renewenvironment{#1}%
   {\linenomath\csname old#1\endcsname}%
   {\csname oldend#1\endcsname\endlinenomath}%
}
\newcommand*\patchBothAmsMathEnvironmentsForLineno[1]{%
  \patchAmsMathEnvironmentForLineno{#1}%
  \patchAmsMathEnvironmentForLineno{#1*}%
}
\AtBeginDocument{%
\patchBothAmsMathEnvironmentsForLineno{equation}%
\patchBothAmsMathEnvironmentsForLineno{align}%
\patchBothAmsMathEnvironmentsForLineno{flalign}%
\patchBothAmsMathEnvironmentsForLineno{alignat}%
\patchBothAmsMathEnvironmentsForLineno{gather}%
\patchBothAmsMathEnvironmentsForLineno{multline}%
\patchBothAmsMathEnvironmentsForLineno{eqnarray}%
}

\usepackage{hyperxmp}

\usepackage[pdftex,
            pdfauthor={\paperauthors},
            pdftitle={\paperasciititle},
            pdfkeywords={\paperkeywords},
            pdfcopyright={Copyright (C) \papercopyright},
            pdflicenseurl={\paperlicenceurl}]{hyperref}

\usepackage[colorinlistoftodos,textsize=scriptsize]{todonotes}

\usepackage[bottom,flushmargin,hang,multiple]{footmisc}

\usepackage[all]{hypcap} 

\usepackage{xspace} 
\usepackage{upgreek}

\def\lhcb   {\mbox{LHCb}\xspace}

\def\MagUp {\mbox{\em Mag\kern -0.05em Up}\xspace}

\ifthenelse{\boolean{uprightparticles}}%
{

 \def\Pmu         {\ensuremath{\upmu}\xspace}                 
 \def\Pnu         {\ensuremath{\upnu}\xspace}                 
                  
 \def\Ppi         {\ensuremath{\uppi}\xspace}

 \def\PDelta      {\ensuremath{\Delta}\xspace}                 
 \def\PXi         {\ensuremath{\Xi}\xspace}                 
 \def\PLambda     {\ensuremath{\Lambda}\xspace}                 
 \def\PSigma      {\ensuremath{\Sigma}\xspace}                 
 \def\POmega      {\ensuremath{\Omega}\xspace}                 
 \def\PUpsilon    {\ensuremath{\Upsilon}\xspace}
 \let\oldPi\Pi
 \def\PPi         {\ensuremath{\oldPi}\xspace}

 \def\PB      {\ensuremath{\mathrm{B}}\xspace}                 
                  
 \def\PD      {\ensuremath{\mathrm{D}}\xspace}

 \def\PK      {\ensuremath{\mathrm{K}}\xspace}

 \def\Pb      {\ensuremath{\mathrm{b}}\xspace}                 
 \def\Pc      {\ensuremath{\mathrm{c}}\xspace}

 \def\Pi      {\ensuremath{\mathrm{i}}\xspace}

 \def\Ps      {\ensuremath{\mathrm{s}}\xspace}

 \def\thebaroffset{0.0em}
}
{

 \def\Pmu         {\ensuremath{\mu}\xspace}                 
 \def\Pnu         {\ensuremath{\nu}\xspace}                 
                  
 \def\Ppi         {\ensuremath{\pi}\xspace}

 \mathchardef\PDelta="7101
 \mathchardef\PXi="7104
 \mathchardef\PLambda="7103
 \mathchardef\PSigma="7106
 \mathchardef\POmega="710A
 \mathchardef\PUpsilon="7107
 \mathchardef\PPi="7105
                  
 \def\PB      {\ensuremath{B}\xspace}                 
                  
 \def\PD      {\ensuremath{D}\xspace}

 \def\PK      {\ensuremath{K}\xspace}

 \def\Pb      {\ensuremath{b}\xspace}                 
 \def\Pc      {\ensuremath{c}\xspace}

 \def\Pi      {\ensuremath{i}\xspace}

 \def\Ps      {\ensuremath{s}\xspace}

 \def\thebaroffset{0.18em}
}
\newcommand{\offsetoverline}[2][\thebaroffset]{\kern #1\overline{\kern -#1 #2}}%

\makeatletter
\ifcase \@ptsize \relax
  \newcommand{\miniscule}{\@setfontsize\miniscule{4}{5}}
\or
  \newcommand{\miniscule}{\@setfontsize\miniscule{5}{6}}
\or
  \newcommand{\miniscule}{\@setfontsize\miniscule{5}{6}}
\fi
\makeatother

\DeclareRobustCommand{\optbar}[1]{\shortstack{{\miniscule (\rule[.5ex]{1.25em}{.18mm})}
  \\ [-.7ex] $#1$}}

\def\mun        {{\ensuremath{\Pmu^-}}\xspace} 

\def\neub       {{\ensuremath{\overline{\Pnu}}}\xspace}

\def\squark    {{\ensuremath{\Ps}}\xspace}

\def\cquark    {{\ensuremath{\Pc}}\xspace}

\def\bquark    {{\ensuremath{\Pb}}\xspace}

\def\pion   {{\ensuremath{\Ppi}}\xspace}

\def\pip    {{\ensuremath{\pion^+}}\xspace}
\def\pim    {{\ensuremath{\pion^-}}\xspace}

\def\kaon    {{\ensuremath{\PK}}\xspace}

\def\KorKbar {\kern \thebaroffset\optbar{\kern -\thebaroffset \PK}{}\xspace}

\def\Kp      {{\ensuremath{\kaon^+}}\xspace}
\def\Km      {{\ensuremath{\kaon^-}}\xspace}

\def\KS      {{\ensuremath{\kaon^0_{\mathrm{S}}}}\xspace}

\def\Kstarp  {{\ensuremath{\kaon^{*+}}}\xspace}
\def\Kstarm  {{\ensuremath{\kaon^{*-}}}\xspace}

\def\Dbar    {{\ensuremath{\offsetoverline{\PD}}}\xspace}
\def\D       {{\ensuremath{\PD}}\xspace}

\def\DorDbar {\kern \thebaroffset\optbar{\kern -\thebaroffset \PD}\xspace}
\def\Dz      {{\ensuremath{\D^0}}\xspace}
\def\Dzb     {{\ensuremath{\Dbar{}^0}}\xspace}
\def\Dp      {{\ensuremath{\D^+}}\xspace}
\def\Dm      {{\ensuremath{\D^-}}\xspace}

\def\DpDm    {\ensuremath{\Dp {\kern -0.16em \Dm}}\xspace}

\def\Dstarp  {{\ensuremath{\D^{*+}}}\xspace}

\def\Ds      {{\ensuremath{\D^+_\squark}}\xspace}
\def\Dsp     {{\ensuremath{\D^+_\squark}}\xspace}

\def\B       {{\ensuremath{\PB}}\xspace}
\def\Bbar    {{\ensuremath{\offsetoverline{\PB}}}\xspace}

\def\BorBbar {\kern \thebaroffset\optbar{\kern -\thebaroffset \PB}\xspace}
\def\Bz      {{\ensuremath{\B^0}}\xspace}
\def\Bzb     {{\ensuremath{\Bbar{}^0}}\xspace}
\def\Bd      {{\ensuremath{\B^0}}\xspace}

\def\BdorBdbar {\kern \thebaroffset\optbar{\kern -\thebaroffset \Bd}\xspace}

\def\Bub     {{\ensuremath{\B^-}}\xspace}

\def\Bpm     {{\ensuremath{\B^\pm}}\xspace}

\def\Bs      {{\ensuremath{\B^0_\squark}}\xspace}
\def\Bsb     {{\ensuremath{\Bbar{}^0_\squark}}\xspace}
\def\BsorBsbar {\kern \thebaroffset\optbar{\kern -\thebaroffset \Bs}\xspace}

\def\Y#1S{\ensuremath{\PUpsilon{(#1S)}}\xspace}

\def\LorLbar     {\kern \thebaroffset\optbar{\kern -\thebaroffset \PLambda}\xspace}

\newcommand{\decay}[2]{\ensuremath{#1\!\to #2}\xspace} 

\def\to                 {\ensuremath{\rightarrow}\xspace}

\def\CP                {{\ensuremath{C\!P}}\xspace}

\def\AT#1     {\ensuremath{A_{\mathrm{T}}^{#1}}\xspace}           

\def\C#1      {\ensuremath{\mathcal{C}_{#1}}\xspace}                       
\def\Cp#1     {\ensuremath{\mathcal{C}_{#1}^{'}}\xspace}                    
\def\Ceff#1   {\ensuremath{\mathcal{C}_{#1}^{\mathrm{(eff)}}}\xspace}        
\def\Cpeff#1  {\ensuremath{\mathcal{C}_{#1}^{'\mathrm{(eff)}}}\xspace}       
\def\Ope#1    {\ensuremath{\mathcal{O}_{#1}}\xspace}                       
\def\Opep#1   {\ensuremath{\mathcal{O}_{#1}^{'}}\xspace}                    

\def\ycp        {\ensuremath{y_{\CP}}\xspace}

\newcommand{\nospaceunit}[1]{\ensuremath{\text{#1}}}       
\newcommand{\aunit}[1]{\ensuremath{\text{\,#1}}}       

\newcommand{\tev}{\aunit{Te\kern -0.1em V}\xspace}
\newcommand{\gev}{\aunit{Ge\kern -0.1em V}\xspace}
\newcommand{\mev}{\aunit{Me\kern -0.1em V}\xspace}
\newcommand{\kev}{\aunit{ke\kern -0.1em V}\xspace}
\newcommand{\ev}{\aunit{e\kern -0.1em V}\xspace}
 
\newcommand{\mevc}{\ensuremath{\aunit{Me\kern -0.1em V\!/}c}\xspace}
\newcommand{\gevc}{\ensuremath{\aunit{Ge\kern -0.1em V\!/}c}\xspace}
\newcommand{\mevcc}{\ensuremath{\aunit{Me\kern -0.1em V\!/}c^2}\xspace}
\newcommand{\gevcc}{\ensuremath{\aunit{Ge\kern -0.1em V\!/}c^2}\xspace}

\def\mum  {\ensuremath{\,\upmu\nospaceunit{m}}\xspace}

\def\fb   {\ensuremath{\aunit{fb}}\xspace}
\def\invfb   {\ensuremath{\fb^{-1}}\xspace}

\newcommand{\stat}{\aunit{(stat)}\xspace}
\newcommand{\syst}{\aunit{(syst)}\xspace}

\newcommand{\chisq}{\ensuremath{\chi^2}\xspace}

\def\gsim{{~\raise.15em\hbox{$>$}\kern-.85em
          \lower.35em\hbox{$\sim$}~}\xspace}
\def\lsim{{~\raise.15em\hbox{$<$}\kern-.85em
          \lower.35em\hbox{$\sim$}~}\xspace}

\def\rad{\aunit{rad}\xspace}

\def\evtgen     {\mbox{\textsc{EvtGen}}\xspace}

\def\geant      {\mbox{\textsc{Geant4}}\xspace}

\def\photos     {\mbox{\textsc{Photos}}\xspace}

\def\pythia     {\mbox{\textsc{Pythia}}\xspace}

\def\tell1  {TELL1\xspace}
\def\ukl1   {UKL1\xspace}

\newcommand{\lhcborcid}[1]{\href{https://orcid.org/#1}{\hspace*{0.1em}\raisebox{-0.45ex}{\includegraphics[width=1em]{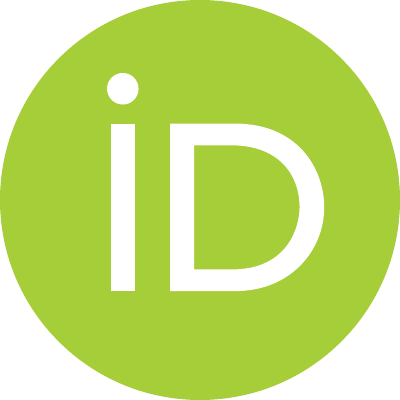}}}}

\usepackage{cite} 
\usepackage{mciteplus}
\usepackage{lipsum}
\usepackage{comment}
\usepackage{subfig}

\newcommand{\PromptDecay}{\ensuremath{\Dstarp\to\Dz(\to\KS\pip\pim)\pip}\xspace}
\newcommand{\SLDecay}{\ensuremath{\Bbar\to\Dz(\to\KS\pip\pim)\mun \bar{\nu}_\mu X}\xspace}

\newcommand{\BdDecay}{\ensuremath{\Bzb\to\Dstarp(\to\Dz\pip)\mun \neub_{\mu} X}\xspace}

\newcommand{\kspp}{\ensuremath{\KS\pip\pim}\xspace}
\newcommand{\Dkspp}{\ensuremath{\Dz\to\KS\pip\pim}\xspace}
\newcommand{\Dkp}{\ensuremath{\Dz\to\Km\pip}\xspace}

\newcommand{\BuDkp}{\ensuremath{\Bub\to\Dz(\to \Km \pip)\mun X}\xspace}

\newcommand{\Dsphipi}{\ensuremath{\Ds\to\phi(\to \Kp \Km)\pip}\xspace}
\newcommand{\Dspipipi}{\ensuremath{\Ds\to\pip\pip\pim}\xspace}
\newcommand{\phiKK}{\ensuremath{\phi\to \Kp \Km}\xspace}

\newcommand{\mpp}{\ensuremath{m^2(\pip\pim)}\xspace}

\newcommand{\cospp}{\ensuremath{|\cos\theta_{\pip\pim}|}\xspace}
\newcommand{\vcospp}{\ensuremath{\cos\theta_{\pip\pim}}\xspace}

\newcommand{\zcp}{\ensuremath{z_{\CP}}\xspace}
\newcommand{\xcp}{\ensuremath{x_{\CP}}\xspace}
\renewcommand{\ycp}{\ensuremath{y_{\CP}}\xspace}
\newcommand{\deltaz}{\ensuremath{\Delta z}\xspace}
\newcommand{\deltax}{\ensuremath{\Delta x}\xspace}
\newcommand{\deltay}{\ensuremath{\Delta y}\xspace}

\newcommand{\re}[2][()] {\ifthenelse{\equal{#1}{()}}{{\ensuremath{{\rm \, Re}}\left(#2\right)}}
                                                    {{\ensuremath{{\rm \, Re}}\left[#2\right]}}}
\newcommand{\im}[2][()] {\ifthenelse{\equal{#1}{()}}{{\ensuremath{{\rm \, Im}}\left(#2\right)}}
                                                    {{\ensuremath{{\rm \, Im}}\left[#2\right]}}}

\newcommand{\gpsqcp}{\displaystyle 1 + \frac{1}{4}\,\langle t^2\rangle_j\,\re{\zcp^2-\deltaz^2}}
\newcommand{\gpsqcprb}{\displaystyle r_b + \frac{1}{4}\,r_b\,\langle t^2\rangle_j\,\re{\zcp^2-\deltaz^2}}

\newcommand{\timebinsSL}{\ensuremath{[0.00, 0.155, 0.285, 0.42, 0.57, 0.74, 0.94, 1.20, 1.58, 2.22, 20.00]\,\tau_\Dz}\xspace}

\newcommand{\xUnits}{\times10^{-3}}
\newcommand{\yUnits}{\times10^{-3}}
\newcommand{\dxUnits}{\times10^{-3}}
\newcommand{\dyUnits}{\times10^{-3}}

\usepackage{booktabs}
\usepackage{longtable} 
\usepackage{multirow}

\begin{document}

\renewcommand{\thefootnote}{\fnsymbol{footnote}}
\setcounter{footnote}{1}


\begin{titlepage}
\pagenumbering{roman}

\vspace*{-1.5cm}
\centerline{\large EUROPEAN ORGANIZATION FOR NUCLEAR RESEARCH (CERN)}
\vspace*{1.5cm}
\noindent
\begin{tabular*}{\linewidth}{lc@{\extracolsep{\fill}}r@{\extracolsep{0pt}}}
\ifthenelse{\boolean{pdflatex}}%
{\vspace*{-1.5cm}\mbox{\!\!\!\includegraphics[width=.14\textwidth]{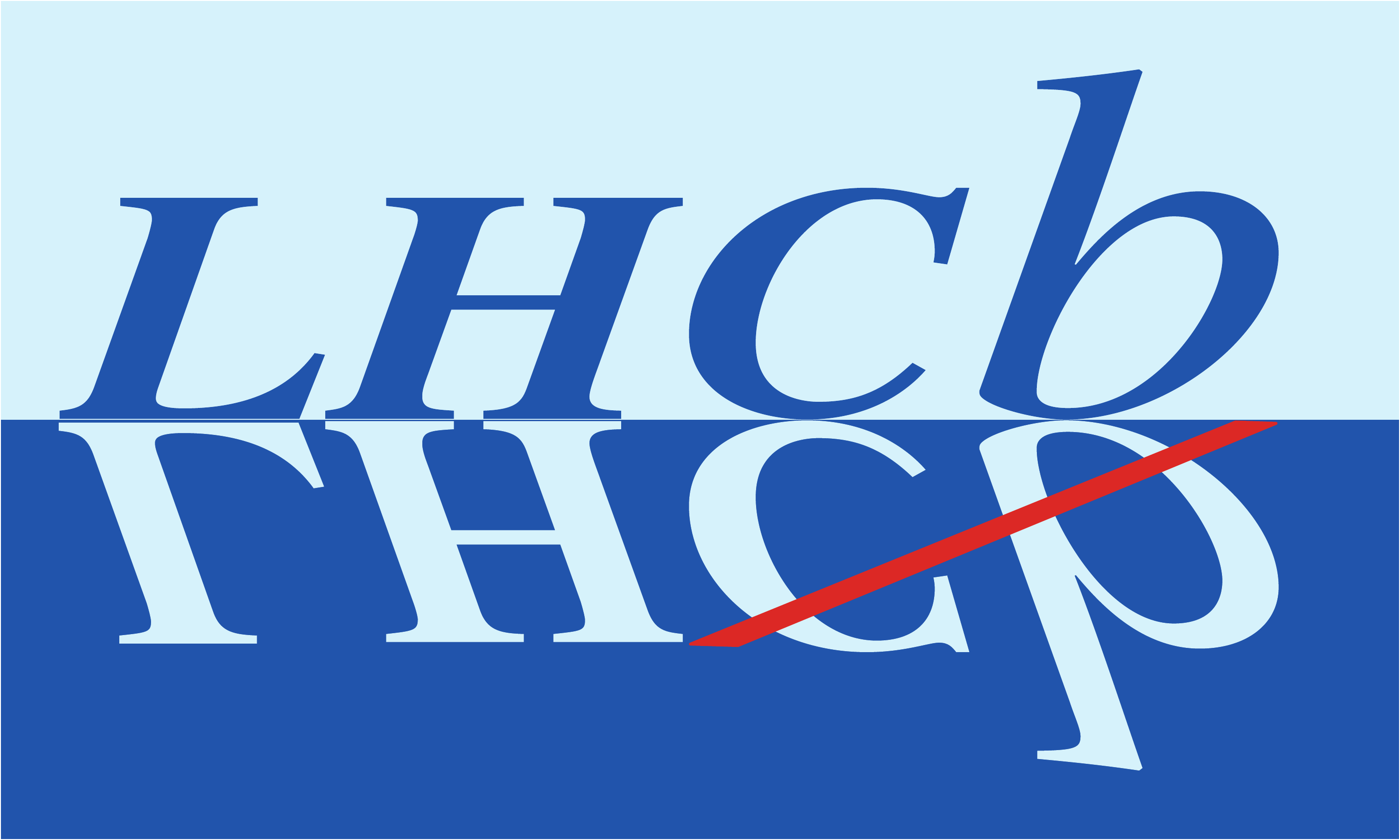}} & &}%
{\vspace*{-1.2cm}\mbox{\!\!\!\includegraphics[width=.12\textwidth]{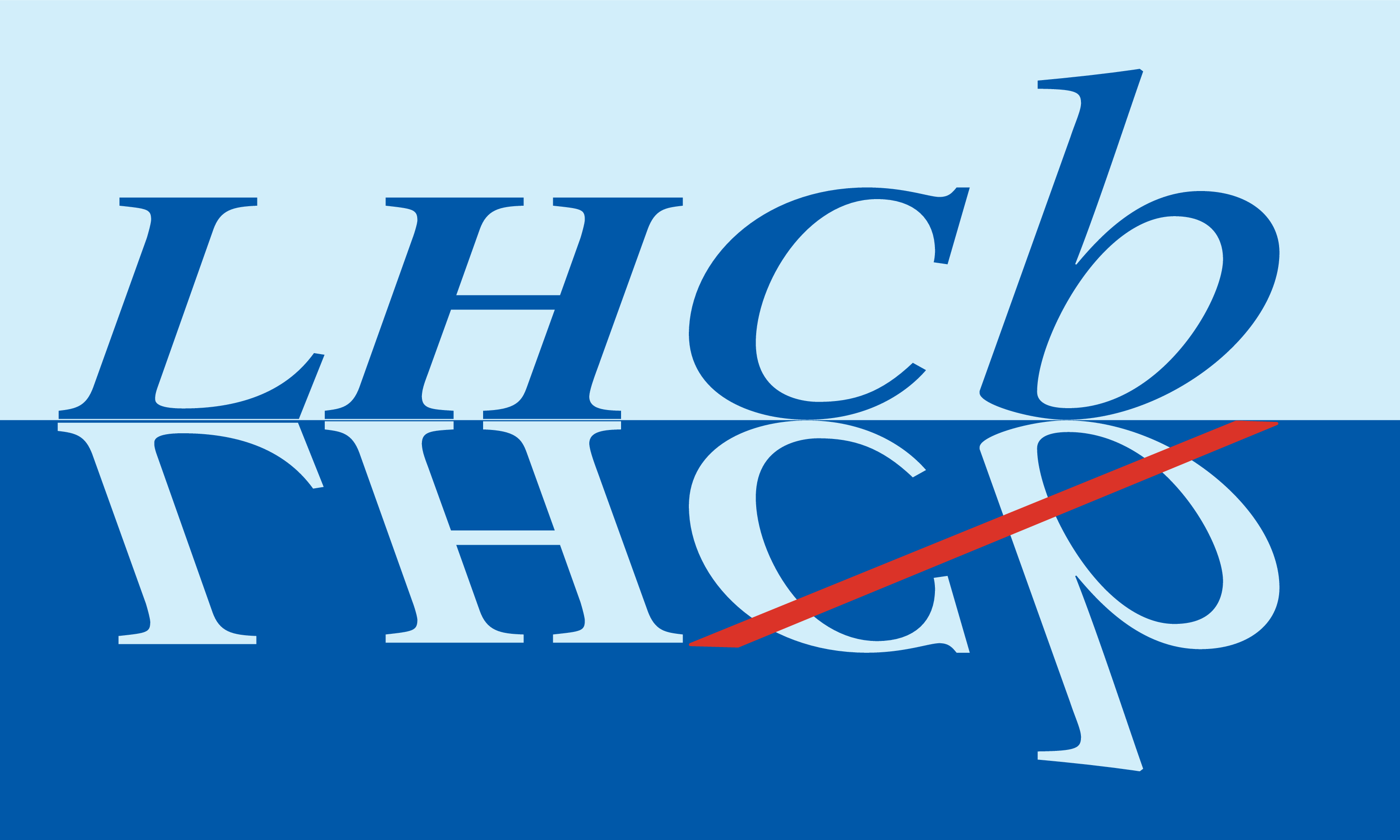}} & &}%
\\
 & & CERN-EP-2022-160 \\  
 & & LHCb-PAPER-2022-020 \\  
 & & \today \\ 
 & & \\
\end{tabular*}

\vspace*{2.0cm}

{\normalfont\bfseries\boldmath\huge
\begin{center}
  \papertitle 
\end{center}
}

\vspace*{1.5cm}

\begin{center}

\paperauthors \footnote{Authors are listed at the end of this paper.}
\end{center}

\vspace{\fill}

\begin{abstract}
  \noindent
     A measurement of charm mixing and \CP-violating parameters is reported, using \SLDecay decays reconstructed in proton-proton collisions collected by the LHCb experiment during the years 2016 to 2018, corresponding to an integrated luminosity of 5.4\invfb.
     The measured mixing and \CP-violating parameters are
     \begin{align*}
         \xcp    &= [  \,\,\,\,\,4.29 \pm  1.48  \stat  \pm  0.26  \syst  ]\xUnits \,, \\
         \ycp    &= [ \,\,12.61 \pm  3.12  \stat  \pm  0.83  \syst  ]\yUnits \,, \\
         \deltax &= [ -0.77 \pm  0.93  \stat  \pm  0.28  \syst  ]\dxUnits \,,\\ 
         \deltay &= [  \,\,\,\,\,3.01 \pm  1.92  \stat  \pm  0.26  \syst  ]\dyUnits \,.
     \end{align*}
     The results are complementary to and consistent with previous measurements. A combination with the recent LHCb analysis of \PromptDecay decays is reported.
\end{abstract}

\vspace*{2.0cm}

\begin{center}
  Published in
    \href{https://journals.aps.org/prd/abstract/10.1103/PhysRevD.108.052005}{Phys.~Rev.~D108 (2023) 052005} 
\end{center}

\vspace{\fill}

{\footnotesize 
\centerline{\copyright~\papercopyright. \href{\paperlicenceurl}{\paperlicence}.}}
\vspace*{2mm}

\end{titlepage}


\newpage
\setcounter{page}{2}
\mbox{~}


\renewcommand{\thefootnote}{\arabic{footnote}}
\setcounter{footnote}{0}

\cleardoublepage


\pagestyle{plain} 
\setcounter{page}{1}
\pagenumbering{arabic}


\section{INTRODUCTION}
\label{sec:introduction}

Flavor oscillation is the transition between a neutral flavored meson and its antiparticle. 
In the Standard Model (SM) of particle physics, this transition is mediated by charged-current weak interactions, involving the exchange of two virtual $W$ bosons. A contribution from unknown massive virtual particles could interfere with the SM oscillation amplitude. This phenomenon is hence sensitive to physics beyond the SM at large scales~\cite{ARNPS.012809.104534}.

The oscillation occurs because the quark mass terms in the SM Lagrangian cannot be simultaneously diagonalized with the weak coupling terms. 
The mass eigenstates of the neutral charm meson can be written as linear combinations of the flavor eigenstates as
$|D_{1,2}\rangle = p |\Dz\rangle \pm q |\Dzb\rangle,$
where $p$ and $q$ are complex parameters satisfying the normalization condition $|p|^2 + |q|^2 = 1$. 
The quantities $m_{1,2}$ and $\Gamma_{1,2}$ are the mass and decay width of the $D_{1,2}$ states, respectively.
The oscillation can be described with two dimensionless parameters,
\begin{alignat}{3}
    x &= (m_1-m_2)c^2/\Gamma \,,\\
    y &= (\Gamma_1 - \Gamma_2) / (2\Gamma)\,, \label{eq:y-def}
\end{alignat}
where $\Gamma=(\Gamma_1+\Gamma_2)/2$ is the average decay width.

In this formalism, \CP violation in mixing can manifest itself through a deviation of $|q/p|$ from unity. If the \Dz and \Dzb mesons decay to a common final state $f$, a nonzero phase $\phi_f \equiv \mathrm{arg}(q\bar{A}_f/pA_f)$ can arise from \CP violation in the interference between mixing and decay. Here, $A_f$ ($\bar{A}_f$) denotes the $\Dz\to f$ ($\Dzb\to f$) decay amplitude. If \CP symmetry is conserved in the decay amplitude,  the \CP-violating phase is final-state independent and denoted as $\phi$.

The parameters of interest are expressed in terms of the \CP-averaged mixing parameters
\begin{alignat}{2}
    \xcp &= \frac{1}{2}\left[x\cos\phi\left(\left|\frac{q}{p}\right|+\left|\frac{p}{q}\right|\right)+y\sin\phi\left(\left|\frac{q}{p}\right|-\left|\frac{p}{q}\right|\right)\right],\label{eq:xcp-def}\\
    \ycp &= \frac{1}{2}\left[y\cos\phi\left(\left|\frac{q}{p}\right|+\left|\frac{p}{q}\right|\right)-x\sin\phi\left(\left|\frac{q}{p}\right|-\left|\frac{p}{q}\right|\right)\right], \label{eq:ycp-def}
\end{alignat}
and the \CP-violating differences
\begin{alignat}{3}
    \deltax &=  \frac{1}{2}\left[x\cos\phi\left(\left|\frac{q}{p}\right|-\left|\frac{p}{q}\right|\right)+y\sin\phi\left(\left|\frac{q}{p}\right|+\left|\frac{p}{q}\right|\right)\right],\label{eq:dx-def}\\
    \deltay &= \frac{1}{2}\left[y\cos\phi\left(\left|\frac{q}{p}\right|-\left|\frac{p}{q}\right|\right)-x\sin\phi\left(\left|\frac{q}{p}\right|+\left|\frac{p}{q}\right|\right)\right]. \label{eq:dy-def}
\end{alignat}
Absence of \CP violation ($|q/p|=1$, $\phi=0$) implies $\xcp=x$, $\ycp=y$, and $\deltax=\deltay=0$.

Oscillations in the $K$ and $B$ meson systems are well established~\cite{PhysRev.103.1901,ALBRECHT1987245,PhysRevLett.97.242003,PDG2020}. 
The evidence and observation of \Dz--\Dzb oscillations were reported much later by the \emph{BABAR} \cite{PhysRevLett.98.211802}, Belle \cite{PhysRevLett.98.211803} and LHCb \cite{LHCb-PAPER-2012-038} collaborations, because of the small oscillation probability governed by the size of the $x$ and $y$ parameters. 
The value of $x$ has only recently been  measured to significantly differ from zero~\cite{LHCb-PAPER-2021-009}. Moreover, \CP violation in the charm sector has been experimentally confirmed much later than in the $K$ and $B$ meson systems. To date, only a single measurement with significance greater than $5\sigma$ exists~\cite{LHCb-PAPER-2019-006} for the difference in time-integrated \CP violation in \decay{\Dz}{\Km\Kp} and \decay{\Dz}{\pim\pip} decays.\footnote{Charge conjugation is implied throughout the paper.}
There have been no experimental indications of \CP violation in mixing or in the interference between mixing and decay of neutral charm mesons thus far. 
The current world averages of mixing and \CP-violating parameters are~\cite{HFLAV18}
\begin{align*}
         x    &= (0.409\,^{+\,0.048}_{-\,0.049})\times10^{-2} \,, \\
         y    &= (0.615\,^{+\,0.056}_{-\,0.055})\times10^{-2} \,, \\
         |q/p| &= \phantom{-}0.995 \pm 0.016 \,,\\ 
         \phi &= -0.044 \pm 0.021 \,\rad.
\end{align*}

The self-conjugate decay \Dkspp provides direct access to both the charm mixing and \CP-violating parameters. 
Using this decay, with the \Dz produced in the decay chain \PromptDecay, the LHCb collaboration reported the first observation of a nonzero value for the $x$ parameter~\cite{LHCb-PAPER-2021-009}.

This paper presents a measurement of charm mixing parameters in \Dkspp decays reconstructed in proton-proton ($pp$) collision data, collected by the LHCb experiment between 2016 and 2018 (Run~2), corresponding to an integrated luminosity of 5.4\invfb. The \Dz mesons originate from semileptonic decays of $b$ hadrons of the form  \SLDecay, where the \Dz flavor is determined from the charge of the muon. The measurement is based on the so-called {\em bin-flip} approach, 
an improved model-independent method, based on~\cite{Bondar:2010qs,Thomas:2012qf}, that
suppresses biases due to a nonuniform event reconstruction efficiency
as a function of phase space and decay-time~\cite{PhysRevD.99.012007}.
This measurement complements the above-mentioned analysis of \PromptDecay decays~\cite{LHCb-PAPER-2021-009}. The independent data sample of \Dz mesons from semileptonic decays allows to sample the low decay-time region, which is not accessible to the \PromptDecay decays analysis.
The procedure of the analysis presented here is mostly aligned with that reported in Ref.~\cite{LHCb-PAPER-2021-009}. A combination of the two results is performed to exploit the increased data sample size and wider coverage of \Dz decay-time.

\section{THE LHCB DETECTOR}
\label{sec:detector}

The LHCb detector~\cite{LHCb-DP-2008-001,LHCb-DP-2014-002} is a single-arm forward spectrometer covering the pseudorapidity range $2 < \eta < 5$, designed for the study of particles containing \bquark\ or \cquark\ quarks. The detector includes a high-precision tracking system consisting of a silicon-strip vertex detector surrounding the $pp$ interaction region, a large-area silicon-strip detector located upstream of a dipole magnet with a bending power of about 4 Tm, and three stations of silicon-strip detectors and straw drift tubes placed downstream of the magnet. The tracking system provides a measurement of the momentum, $p$, of charged particles with a relative uncertainty that varies from 0.5\% at low momentum to 1.0\% at
200\gevc. The minimum distance of a track to a primary $pp$ collision vertex (PV), the impact parameter (IP), is measured with a resolution of $(15 + 29/p_{\mathrm{T}}$)\,$\mum$, where $p_{\mathrm{T}}$ is the component of the momentum transverse to the beam, in \gevc. Different types of charged hadrons are distinguished using information from two ring-imaging Cherenkov
detectors.
Photons, electrons and hadrons are identified by a calorimeter system consisting of scintillating-pad and preshower detectors, an electromagnetic and a hadronic calorimeter. Muons are identified by a system composed of alternating layers of iron and multiwire proportional chambers. The online event selection is performed by a trigger, which consists of a hardware stage, based on information from the calorimeter and muon systems, followed by a software stage, which applies a full event reconstruction.

Simulation is required to model the effects of the detector acceptance and the
  imposed selection requirements.
  In the simulation, $pp$ collisions are generated using
  \pythia~\cite{Sjostrand:2007gs,*Sjostrand:2006za} 
  with a specific \lhcb configuration~\cite{LHCb-PROC-2010-056}.
  Decays of unstable particles
  are described by \evtgen~\cite{Lange:2001uf}, in which final-state
  radiation is generated using \photos~\cite{davidson2015photos}.
  The interaction of the generated particles with the detector, and its response,
  are implemented using the \geant
  toolkit~\cite{Allison:2006ve, *Agostinelli:2002hh} as described in
  Ref.~\cite{LHCb-PROC-2011-006}.

\section{ANALYSIS METHOD}
\label{sec:ANAstrategy}

The analysis is based on the bin-flip method proposed in Ref.~\cite{PhysRevD.99.012007}. It is a model-independent approach, optimized for  the measurement of the mixing parameter $x$, which avoids the need for an accurate modeling of the efficiency variation across phase space and decay-time. The relevant aspects of the method are summarized below.

The \Dkspp decay dynamics is embodied in a Dalitz plot, parametrized with the squared two-body masses,
\begin{equation}
m_\pm^2 \equiv \left\{\begin{aligned}
m^2(\KS\pi^\pm)\quad &\text{for}\ \Dz \to\kspp\ \text{decays}\\
m^2(\KS\pi^\mp)\quad &\text{for}\ \Dzb\to\kspp\ \text{decays} \\
\end{aligned}\right. ~.
\end{equation}
 The parameters of interest are obtained from time-dependent ratios of yields in bins symmetric with respect to the principal bisector of the Dalitz plot, which is defined by $m_+^2 = m_-^2$. The region defined by $m_+^2 > m_-^2$ ($m_+^2 < m_-^2$) is called the lower (upper) region of the Dalitz plot. Among possible intermediate resonances, the \Dz meson decay can proceed through a Cabibbo-favored (CF) path via $\Kstarm\pip$ or a doubly Cabibbo-suppressed (DCS) path via $\Kstarp\pim$.

\begin{figure}
  \begin{center}
    \includegraphics[width=0.48\linewidth]{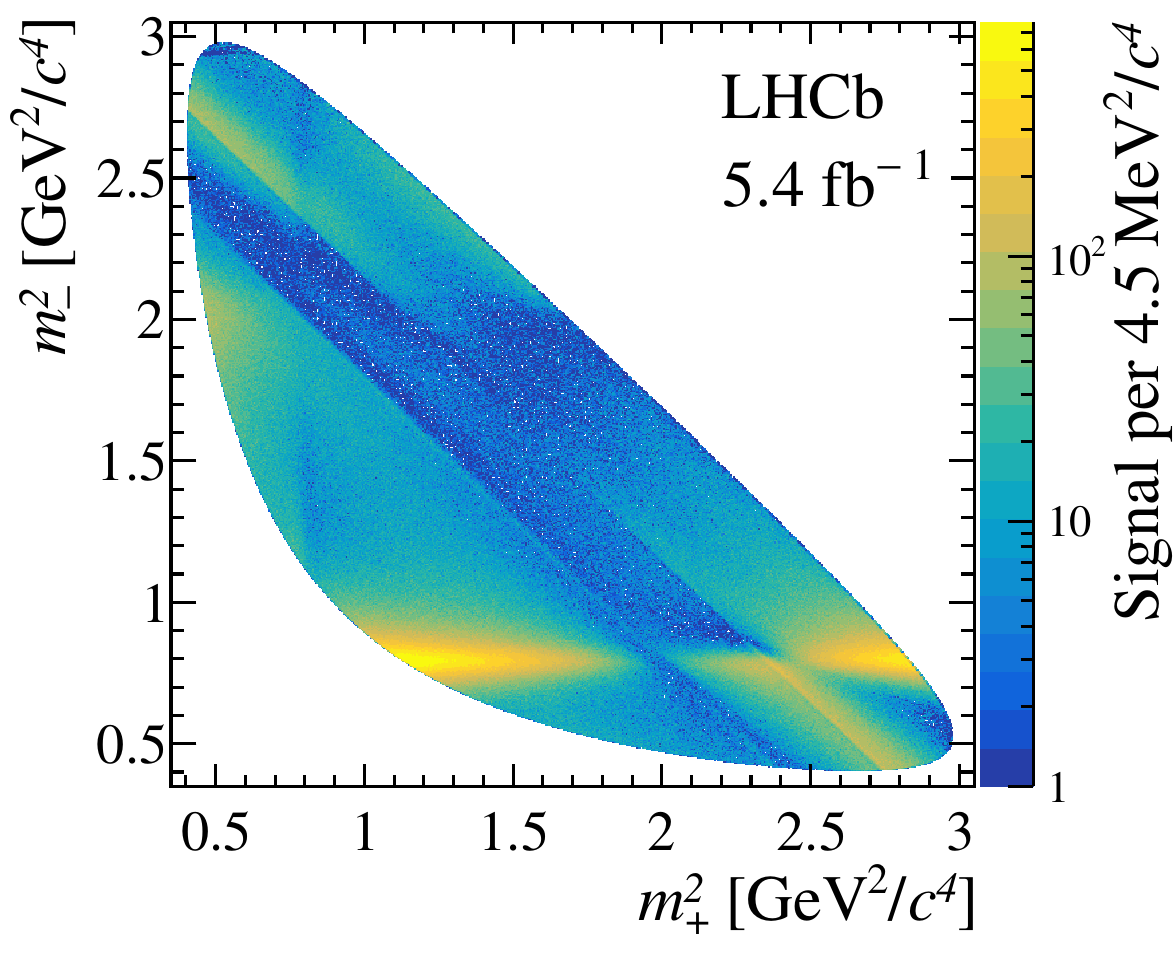}
    \includegraphics[width=0.49\linewidth]{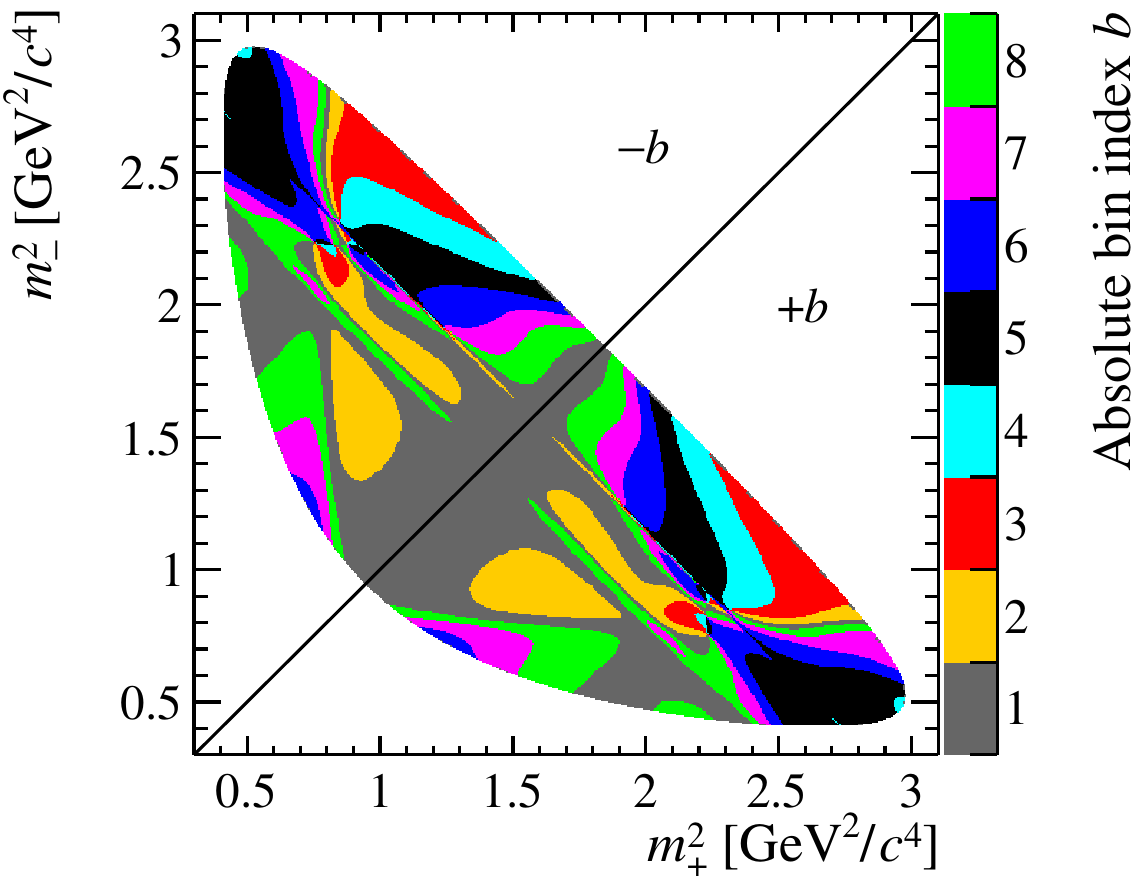}
    \vspace*{-0.5cm}    
  \end{center}
  \caption{
    (Left) \Dz Dalitz plot of reconstructed \SLDecay decays for the \Dz flavor and
   (right) definition of the binning scheme proposed by CLEO~\cite{PhysRevD.82.112006}.
        }
  \label{fig:dalitz-bins}
\end{figure}

These paths populate specific regions in the Dalitz plot, as can be seen in Fig.~\ref{fig:dalitz-bins} (left). The decays proceeding through the CF path dominate in the lower part of the Dalitz plot, while the DCS transitions populate the upper part of the plot. The ratio of decays in these two regions of the Dalitz plot does not change with time in the absence of mixing. In the presence of mixing, the \Dz mesons that have oscillated and decay via the CF path populate the same region as nonmixed mesons decaying via the DCS path. Measuring the time evolution of the ratio between the yields in those regions gives access to the mixing parameters. Separating the data sample by flavor further allows the measurement of \CP-violating parameters.

The Dalitz space is divided into bins such that each bin $b$ in the lower part of the Dalitz plot has a corresponding bin $-b$ in the upper part of the Dalitz plot. 

A scheme with eight pairs of bins as proposed by CLEO~\cite{PhysRevD.82.112006} is used, where bins are chosen such that the strong phase difference between the \Dz and \Dzb amplitudes is nearly constant in each bin. The binning scheme is depicted in Fig.~\ref{fig:dalitz-bins} (right). The data are further divided into ten equipopulated bins with the following edges in the measured \Dz decay-time,
\begin{equation}
\timebinsSL 
\label{eq:time-bin-boundariesSL} \,,
\end{equation}
where $\tau_\Dz$ is the world-average value of the \Dz  lifetime~\cite{PDG2020}.
The yields are measured for each initial flavor of the \Dz meson, Dalitz bin and decay-time bin.

For small mixing parameters and in the limit of \CP-conserving decay amplitudes, the ratio of yields between the Dalitz bin $-b$ and the Dalitz bin $+b$ in the decay-time bin $j$ can be expressed as~\cite{PhysRevD.99.012007}
\begin{equation}
    R_{bj}^\pm \approx \frac{\gpsqcprb + \dfrac{1}{4}\,\langle t^2\rangle_j\, \left|\zcp\pm\deltaz\right|^2 + \sqrt{r_b} \langle t\rangle_j \re[]{X_b^*(\zcp\pm\deltaz)}}{\gpsqcp+r_b\,\dfrac{1}{4}\,\langle t^2\rangle_j\, \left|\zcp\pm\deltaz\right|^2 + \sqrt{r_b} \langle t\rangle_j \re[]{X_b(\zcp\pm\deltaz)}} \,,
    \label{eqn:ratio}
\end{equation}
where the $+$ ($-$) sign refers to the \Dz (\Dzb) initial flavor.
Here $\langle t\rangle_j$ ($\langle t^2\rangle_j$) is the average of the decay-time (squared) of unmixed decays in units of $\tau_\Dz$, 
$r_b$ is the decay-time independent ratio of signal yields
between bins $-b$ and $+b$, $\zcp\pm\deltaz\equiv-\left(q/p\right)^{\pm1}(y+ix)$, and $X_b\equiv c_b-is_b$, 
where $c_b$ and $s_b$ are the amplitude-weighted averages of the cosine and sine of the strong-phase difference over the Dalitz bin $\pm b$. The mixing and \CP-violating parameters are determined by performing a simultaneous fit of the $R_{bj}^\pm$ expressions  
to the measured yield ratios. Equation (\ref{eqn:ratio}) is valid if time-dependent variations of the Dalitz plot efficiency are negligible. Time-independent efficiency variations in the Dalitz phase space do not affect the extraction of the mixing and \CP-violating parameters, which relate to \zcp and \deltaz as 
\begin{alignat}{4}
    \xcp &= -\im{\zcp} \,, \quad& \deltax &= -\im{\deltaz}\,,\\
    \ycp &= -\re{\zcp} \,, \quad& \deltay &= -\re{\deltaz}\,.
\end{alignat}
The analysis steps are described in the following paragraphs, with references to specific sections given where applicable. Section~\ref{sec:CANselection} explains the initial selection of the data, which includes a multivariate analysis (MVA) dedicated to the suppression of the combinatorial background. The bin-flip method assumes no correlation between decay-time and Dalitz-space coordinates, as it integrates over Dalitz and decay-time bins separately. Experimentally induced correlations, caused by nonuniform selection efficiencies, are removed through a combination of simulation-based and data-driven methods, which is described in detail in Sec.~\ref{sec:Decorrelation}. Validation tests are performed to confirm that any remaining reconstruction and selection effects do not affect the final result and hence do not need to be explicitly accounted for. These studies are later used to construct realistic pseudoexperiment models for the study of systematic uncertainties.

The data, weighted according to the decorrelation method, are split into bins of Dalitz space and decay-time, and separated according to the \Dz meson initial flavor. The data are further categorized as LL or DD depending on whether the \KS meson decay products are reconstructed as long or downstream tracks. Long tracks are reconstructed from hits in both the VELO and the downstream tracking stations. Downstream tracks do not use any hit information from the VELO. The data is not split by data-taking year or magnet polarity, and cross-checks are performed to validate that there is no dependence of the results on data taking conditions.

Unbinned maximum likelihood fits to the invariant-mass distribution of the reconstructed \Dz mesons are performed for each category and bin to extract the signal yields. The fit model includes signal and combinatorial background components. The signal is described by a sum of a Johnson SU function~\cite{johnson} and a bifurcated Gaussian function. The combinatorial background is modeled with a first-order (second-order) Chebyshev polynomial for the \KS LL (DD) category.

Equation (\ref{eqn:ratio}) requires as inputs the averages of $t = \tau/\tau_\Dz$ and $t^2$ in each time bin $j$, where $\tau$ is the proper decay-time and $\tau_\Dz$ the average \Dz lifetime. They are computed as
\begin{equation}
\langle t \rangle_j = \frac{\sum_i w_{i}\;t_{i} }{\sum_i w_{i}}\; \quad \mbox{and}\quad  \langle t^2 \rangle_j  = \frac{\sum_i w_{i}\;t_{i}^2 }{\sum_i w_{i}}\;,
\end{equation}
where the sum is over the selected candidates $i$ in decay-time bin $j$ and in the lower part of the Dalitz plot (as this area is dominated by decays of mesons that did not undergo mixing), and $w_{i}$ is the product of a signal weight obtained from the mass fit using the $sPlot$ method~\cite{Pivk:2004ty} and a weight from the decorrelation procedure.

The mixing and \CP-violating parameters are determined using a least-squares fit of the $R_{bj}^\pm$ expressions of Eq.~(\ref{eqn:ratio})
to the $2\times 2\times 8\times 10\times 2 = 640$ measured yields $N^\pm_{\pm bjk}$ and their uncertainties $\sigma^\pm_{\pm bjk}$ in all bins and categories. The  $\chi^2$ function
\begin{align}
\chisq &= \sum_b^8\sum_j^{10}\sum_{k=\rm LL,\,DD}\left[\frac{(N^+_{-bjk}-N^+_{+bjk}R^+_{bj})^2}{(\sigma^+_{-bjk})^2+(\sigma^+_{+bjk}R_{bj}^+)^2}+\frac{(N^-_{-bjk}-N^-_{+bjk}R^-_{bj})^2}{(\sigma^-_{-bjk})^2+(\sigma^-_{+bjk}R_{bj}^-)^2}\right] + \chisq_{X} \label{eq:fit-chi2}
\end{align}
is minimized, where the Gaussian penalty term
\begin{align}
\chisq_{X} &= \sum_b^8\sum_{b'}^8\left(X^{\rm{ext}}_b-X_b\right)(V_{\rm{ext}}^{-1})_{bb'}\left(X^{\rm{ext}}_{b'}-X_{b'}\right) \label{eqn:fit-penalty} 
\end{align}
represents external constraints 
on the eight complex quantities $X_b$ from the combined determinations 
$X_b^{\rm{ext}}$ (with statistical and systematic covariance matrix $V_{\rm{ext}}$)
of the CLEO and BESIII measurements~\cite{PhysRevD.101.112002}.

The free parameters of the $\chi^2$ minimization are \xcp, \ycp, \deltax, \deltay, and the eight ratios $r_b$. 
The results are presented in Sec.~\ref{sec:Result}.

The systematic uncertainties are discussed in Sec.~\ref{sec:SystematicUncertainties}.  They are determined using generated pseudoexperiments that apply reconstruction and selection effects to an amplitude model in order to obtain a realistic description of the data.

This analysis complements the measurement of mixing and \CP-violating parameters with the bin-flip method using \Dz mesons from the decay chain ${\PromptDecay}$~\cite{LHCb-PAPER-2021-009}. Section~\ref{sec:Combination} presents a combination of the two sets of results, which is done by performing a simultaneous fit to the two samples.

\section{CANDIDATE SELECTION}
\label{sec:CANselection}

Candidates are reconstructed in the decay chain \SLDecay, where $X$ represents possible additional decay products that are not reconstructed. The \KS candidates are reconstructed from two oppositely charged pion tracks either in the LL or DD category (see Sec.~\ref{sec:ANAstrategy}). At least one displaced, high-transverse-momentum muon is required. An MVA algorithm is used to select candidates conforming to a topology of an $n$-body ($n = 2, 3, 4$) decay of a $b$-hadron, where at least one of the tracks must be a muon. A fit~\cite{Hulsbergen:2005pu} is then performed on the selected candidates, constraining the \Dz decay tracks to a common origin vertex and the \KS mass to its world average value~\cite{PDG2020}. The analysis uses the reconstructed \Dz mass $m(\KS\pip\pim)$ and decay-time variables obtained from this fit. The Dalitz-plot coordinates are determined in another fit in which the \Dz mass is constrained to its known value~\cite{PDG2020}.

Combinatorial background is further suppressed with a dedicated MVA, namely a boosted decision tree (BDT) classifier~\cite{Friedman:2001wbq}. The signal and background distributions for the training are obtained through the $sPlot$ technique using the fit to the $m(\KS\pip\pim)$ distribution. The signal range is $[1795, 1935]\mevcc$. The BDT employs topological and kinematic variables related to the reconstructed $b$ hadron: the quality of the primary and secondary vertices, the difference in the vertex-fit $\chi^2$ of the PV reconstructed with and without the decay products of the $b$-hadron, the flight distance, the cosine of the angle between the momentum and the vector connecting the primary and secondary vertices, and the corrected mass
$m_{\text{corr}} = \sqrt{m^2(\Dz\mu^-) + p^2_\perp(\Dz\mu^-) } + p_\perp(\Dz\mu^-)$, where $m(\Dz\mu^-)$ is the invariant mass of the $\Dz\mu^-$ combination and $p_\perp(\Dz\mu^-)$ is the component of its momentum perpendicular to the $b$ hadron flight direction.
The only variables related to the $b$-hadron children are the transverse momenta of the muon and the reconstructed \Dz meson. A gradient boosting algorithm is used with uniform regularization (uBoost)~\cite{Stevens_2013}. A manual 6-fold cross-validation is implemented~\cite{crossValidation1999}.

A requirement on the BDT output variable is optimized by maximising the signal significance, defined as $N_{\rm sig}/\sqrt{N_{\rm sig}+N_{\rm bkg}}$, where $N_{\rm sig}$ and $N_{\rm bkg}$ are the numbers of signal and background events, obtained from fits to the \Dz mass distribution. The optimal points are computed separately for each \KS category. Validation tests support the strategy of using a single value for all data-taking years. The selection retains 80\% (73\%) of the signal candidates and increases the signal purity from 26\% (21\%) to 79\% (59\%) for the LL (DD) sample.
\begin{figure}[tb]
  \begin{center}
    \includegraphics[width=0.49\linewidth]{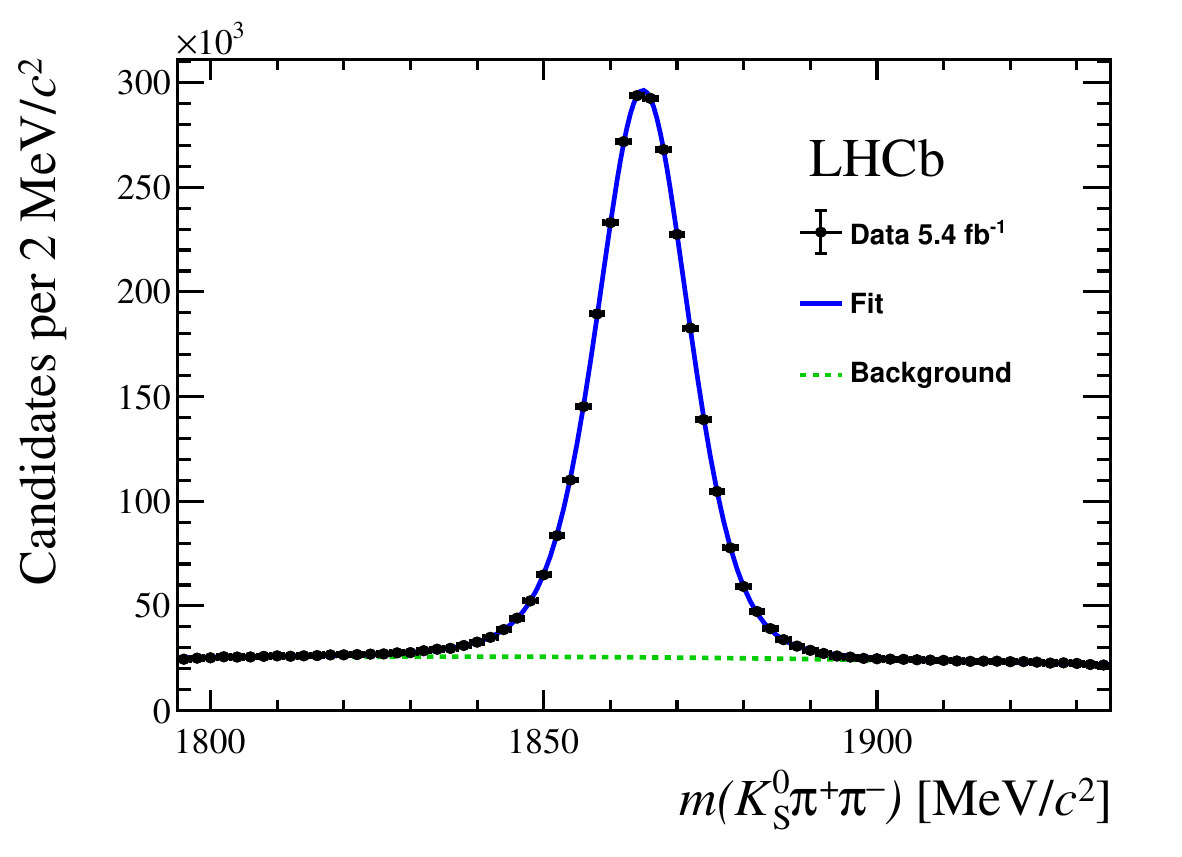} 
    \includegraphics[width=0.49\linewidth]{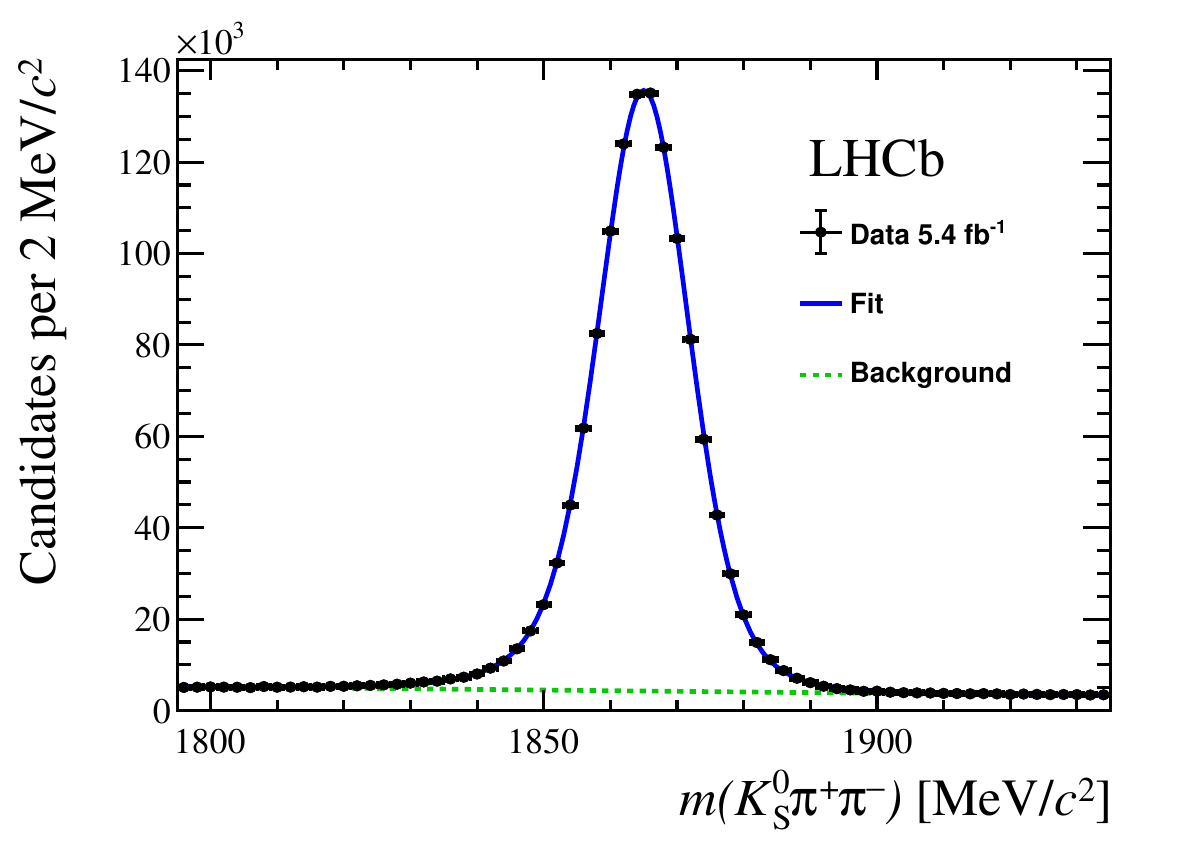}
    \vspace*{-0.5cm}
  \end{center}
  \caption{
   \KS\pip\pim invariant-mass distribution of (left) DD and (right) LL \KS candidates after all selection requirements, with fit results superimposed.
        }
  \label{fig:md-selection}
\end{figure}
Duplicated tracks that originate from the same physical particle are removed for the LL sample by rejecting candidates using tracks for which the slope in the VELO is too similar to that of another track in the event. For the DD sample, in addition to the requirement on the track slopes, a large enough difference in the reconstructed momentum is required for any two tracks. If there are multiple candidates in the event after the clone tracks removal, a single randomly chosen candidate is retained. 

The yields are extracted through a fit to $m(\KS\pip\pim)$ distribution.
The total signal yields after the full selection are 1.24$\times10^6$ (2.48$\times10^6$) for the LL (DD) sample. The \KS\pip\pim invariant-mass distribution after the selection is shown in Fig.~\ref{fig:md-selection}, with fit results superimposed.

\section{DECORRELATION}
\label{sec:Decorrelation}

An important assumption of the analysis method is that there are no experimentally induced correlations between phase-space coordinates and the \Dz candidate decay-time, such that it is possible to integrate separately over the Dalitz bins and in bins of decay-time to obtain the decay-time-independent coefficients $r_b$ and $X_b$. 

\begin{figure}[tb]
  \begin{center}
    \includegraphics[width=0.49\linewidth]{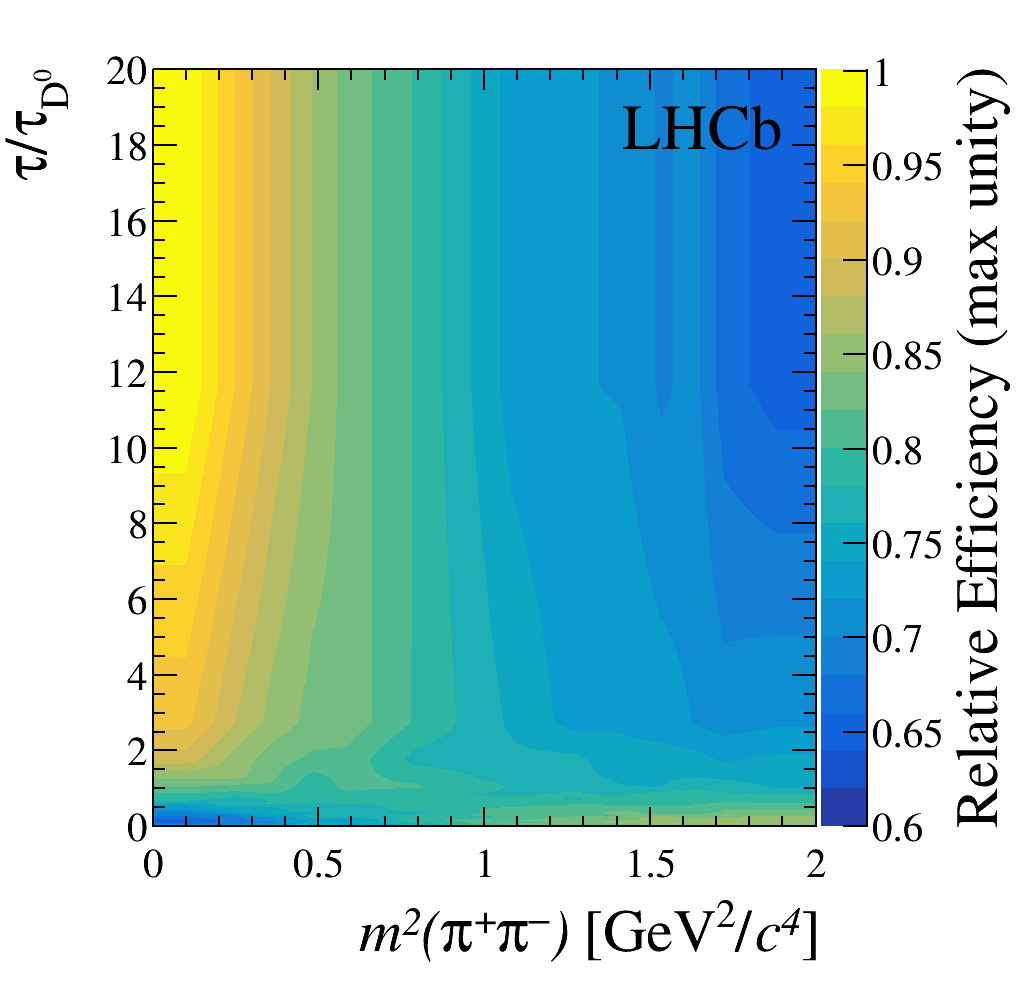}
    \includegraphics[width=0.49\linewidth]{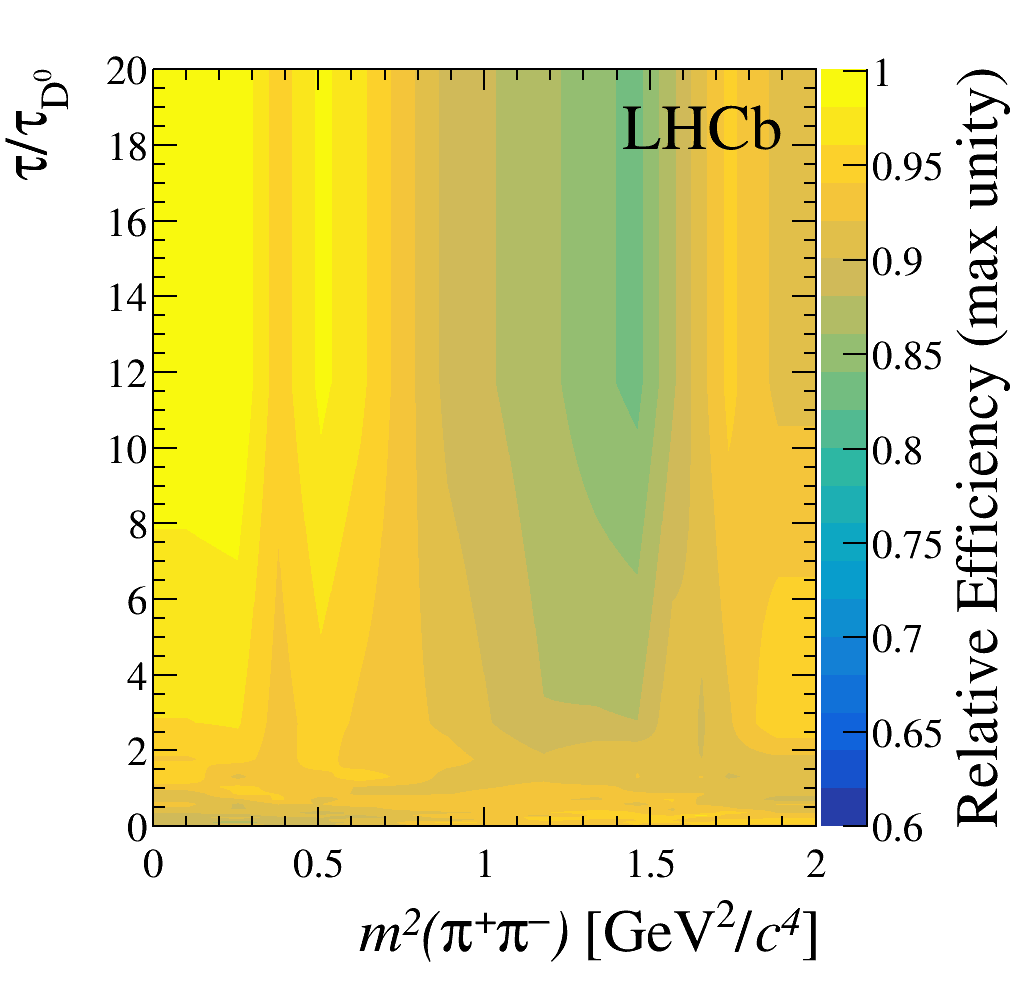}
    \vspace*{-0.5cm}
  \end{center}
  \caption{
    Relative efficiency as a function of \Dz decay-time and \mpp as determined from \SLDecay candidates, 
    separately for (left) DD and (right) LL $\KS$ candidates. The distributions are smoothed using bilinear interpolation.
        }
  \label{fig:correlation}
\end{figure}

Such a correlation has already been observed in the analysis of the decay ${\PromptDecay}$~\cite{LHCb-PAPER-2019-001,LHCb-PAPER-2021-009}. 
In this data sample, it is induced mainly by  the online software selection for $n$-body hadronic $b$ hadron decays ($n = 2,3,4$)~\cite{LHCb-PROC-2015-018}. 
The selection requires that 2, 3 or 4 tracks form a single displaced vertex. In the case of \SLDecay, one of the tracks is required to be a muon, and all other tracks must come from the children of the \Dz meson. Hence, the selection favors configurations where the \Dz meson decays close to the \B-meson vertex, introducing a correlation between the Dalitz coordinates and the \Dz decay-time. 
This is shown in Fig.~\ref{fig:correlation}, where the dependence of the squared invariant mass of the two final-state pions, \mpp, on the normalized \Dz meson decay-time $\tau$/$\tau_\Dz$ is reported. The figure shows the signal yields, normalized to the maximum yield, in each \mpp bin as a function of \Dz decay-time. 
Due to the known small values of the mixing parameters~\cite{HFLAV18} and the \Dkspp amplitude model~\cite{Adachi:2018itz}, 
the correlation between \mpp and the \Dz decay-time caused by the mixing effect is negligible at the current sample size. 

Therefore, the correlation observed in Fig.~\ref{fig:correlation} is induced by the online software selection only.

Using \SLDecay simulated events, the efficiency of the online selection is determined as a function of the \Dz decay-time, \mpp and \vcospp, where $\theta_{\pi^+\pi^-}$ is the angle between the direction vector of the
\pip\pim pair in the \Dz meson rest frame and the direction vector of either pion in the \pip\pim rest frame. Note that each \mpp bin comprises events from  both sides of the bisector of the Dalitz plot, thus the effect of mixing, which changes the ratios in Eq. (\ref{eqn:ratio}) as a function of \Dz decay-time, is not present here. 
The efficiency in this phase space is smoothed and the inverse of the efficiency is assigned as a weight to each signal candidate in data. 
This efficiency correction suppresses to a large extent the correlation induced from the online selection effects.

A further decorrelation procedure based on data is applied to remove the small remaining correlation. 
It is determined from the background-subtracted data, once the weights obtained from the simulation have been applied. 
A decorrelation weight is derived as the inverse of the 
relative proportion of signal candidates observed in each \Dz decay-time and \mpp bin as is done in Fig.~\ref{fig:correlation}. 
In this way, a uniform decay-time acceptance is achieved, without knowledge of the absolute efficiency.
Each candidate is thus weighted with the product of the weight from simulation and the data-driven decorrelation weight. From the weighted data sample, yields are extracted for each Dalitz and \Dz decay-time bin, and fitted with the bin-flip method in Sec.~\ref{sec:Result}. The same combined weight is included to generate realistic pseudoexperiments in Sec.~\ref{sec:SystematicUncertainties} to validate the method used in this analysis.
A similar decorrelation procedure has been used by LHCb in Refs.~\cite{LHCb-PAPER-2019-001,LHCb-PAPER-2021-009}.

\section{SYSTEMATIC UNCERTAINTIES}
\label{sec:SystematicUncertainties}

Pseudoexperiments are used to assess the systematic uncertainties and to validate the analysis procedure.
The pseudoexperiments are generated by sampling the decay-time-dependent decay rate using the Belle model~\cite{Adachi:2018itz} to describe the amplitudes at $t=0$. 
The mixing and \CP-violating parameters are included according to the measured values.  
Phase-space and decay-time acceptance effects are modeled on simulated samples. The correlation between the \Dz decay-time and the Dalitz plot coordinates is generated using the inverse of the decorrelation weights determined in Sec.~\ref{sec:Decorrelation}. For each set of pseudoexperiments, these weights are fluctuated within their statistical uncertainties.
The pseudoexperiment data can then be processed in exactly the same manner as the collision data. 
Samples are generated separately for each of the \KS categories, and processed separately up until the last step when all samples are combined to determine the mixing and \CP-violating parameters. 

To determine the systematic uncertainty due to a given source, pseudodata are generated with the effect in question included. Performing the fit on this dataset determines the bias on the results ensuing from this effect, which however includes statistical effects. This is denoted as the default fit. An additional reference fit is performed where the effect in question is not present or has been corrected for. The difference between the two results represents the systematic uncertainty from the source under consideration.

In Sec.~\ref{sect:reco_sel}, the reference fit is performed with the pseudodata generated without any detector effects. For determining the rest of the systematic uncertainties, the reconstruction and selection effects are included in the pseudodata in order to represent the data realistically. To avoid double-counting, in these cases the reference fit is the default fit from Sec.~\ref{sect:reco_sel}.

\subsection{Reconstruction and selection effects}
\label{sect:reco_sel}

As mentioned above, the reconstruction and selection effects are incorporated into the pseudodata using the acceptance and resolution models obtained from simulation, and the weights from the decorrelation method. 

The dominant systematic uncertainties on \ycp are due to the neglected decay-time and Dalitz-coordinate resolutions, as well as efficiency variations.
Some biases arise in \deltax and \deltay due to the correlation between resolutions of the Dalitz coordinates, stemming from the \KS mass constraint. 

In order to have pseudoexperiments which represent the actual data and to be sure to assess the true effect of the different systematic sources under a realistic setting, these effects are included in all the subsequent studies presented in this section. 
To avoid double counting of these effects, this baseline bias is included as a systematic uncertainty once, and then the effect of the additional systematic uncertainty sources is calculated with respect to it. 

\subsection{Detection asymmetries}
The reconstruction efficiency for tracks originating from charged pions varies 
between the positive and negative charges and depends on momentum. 
In the \Dkspp decay, this affects the efficiency across the Dalitz plot with respect to its bisector and introduces an artificial flavor asymmetry between the \Dz and \Dzb mesons. This induces a bias on the measurement of the \CP-violating parameters \deltax and \deltay. 

The asymmetry in the \Dkspp sample is estimated using two Cabibbo-favored \Ds decays: \Dspipipi and \Dsphipi. The \Ds decay channels are selected with requirements as similar as possible to those for the \Dkspp decays.
In the case of the $\Dspipipi$ decay, the uncorrected, measured asymmetry comprises the asymmetry of a pion pair \pip\pim, $A^{\pip\pim}_{\mathrm{det}}$, the single pion detection asymmetry, $A_\mathrm{det}(\pip)$, the \Ds meson production asymmetry, $A_\mathrm{prod}(\Ds)$, and the asymmetry from the online event selection of the \Ds meson, $A_\mathrm{trig}(\Ds)$.  
Similar components appear for the $\Dsphipi$ decay, with the exception of the $A^{\pip\pim}_{\mathrm{det}}$ component. The asymmetry from \Kp\Km is ignored as \phiKK is a self-conjugate decay in which the phase-space of the kaons is identical, thus canceling any reconstruction asymmetry effect. Hence, the asymmetries from these two decays can be expressed to first order as
\vspace{-5mm}

\begin{align}
    A_{\mathrm{meas}} (\Dspipipi) & = &\  & A_\mathrm{prod}(\Ds) + A_\mathrm{trig}(\Ds) + A_\mathrm{det}(\pip) + A^{\pip\pim}_{\mathrm{det}} \,,   \\ 
    A_{\mathrm{meas}} (\Dsphipi)  & = &\                                & A_\mathrm{prod}(\Ds) + A_\mathrm{trig}(\Ds) + A_\mathrm{det}(\pip) \,,
    \label{eqn:DsExpand}
\end{align}
where in the \Dspipipi decay, one of the pion of equal electric charge is paired randomly with the \pim, and the other pion corresponds to the pion in the ${\Dsphipi}$ decay. 
To first approximation, the difference in uncorrected asymmetries between the two \Dsp decay modes is equal to
\vspace{-1mm}

\begin{equation}
    A^{\pip\pim}_\mathrm{det} = A_\mathrm{meas} (\Dspipipi) - A_\mathrm{meas} (\Dsphipi) \quad .
    \label{eqn:DetAsym}
\end{equation}
These quantities vary over phase-space and kinematic distributions of the \Ds meson. 
A gradient boosting reweighting algorithm~\cite{Rogozhnikov:2016bdp} is implemented to equalize the kinematic distributions of the \Ds and \Dz meson samples in each bin of \mpp, \cospp and \Dz decay-time.

The obtained $A^{\pip\pim}_\mathrm{det}$ are independent of the Dalitz region and \Dz decay-time. They are found to be $-0.017 \pm 0.013$ and $-0.010 \pm 0.016$ for DD and LL \KS candidates, thus compatible with zero.
These values are incorporated in the pseudoexperiment data in addition to the baseline configuration to determine the associated systematic uncertainty. 

\subsection{Mass fit model}

The bin-flip method deals with ratios of yields 
between the upper and lower parts of Dalitz plot,
which are kinematically similar. The analysis is therefore robust against the choice of a fit model, which will affect the numerator and denominator of the ratios in the same way. 
However, a possible systematic bias is examined by considering an alternative mass fit model. 

The systematic uncertainty from the mass fit is estimated by changing the signal PDF from the Johnson distribution to a crystal ball function~\cite{Skwarnicki:1986xj}. 
The sensitivity to the choice of the background model is  investigated with an alternative polynomial model.
The joint alternative fit models are implemented in a fit on the same pseudodata as described in Sec.~\ref{sect:reco_sel}. The systematic effect in the final measurement is found to be very small, confirming that the measurement is very robust with regard to the choice of the fit models.

\subsection{Unrelated \boldmath $\Dz\mun$ combinations}
\label{sec:mistag}
The data sample has some contamination from $\Dz\mun$ combinations in which the muon does not come from the decay of the same $b$ hadron as the $\Dz$ candidate. As the flavor of the $\Dz$ meson is identified from the charge of the accompanying muon, combinations with a random muon have a 50\% chance of wrongly tagging the initial flavor of the \Dz meson. Additionally, the decay-time of the \Dz  candidate is wrongly estimated, as it is extrapolated to a wrong production vertex.

The probability of wrongly tagging the flavor of a candidate is determined using the \BuDkp decay channel, where it can be estimated by comparing the sign of the kaon and muon after accounting for mixing effects and contributions from doubly Cabibbo-suppressed decays~\cite{LHCb-PAPER-2017-046}. It is further calibrated using so-called doubly tagged samples of both \Dkp and signal decays. Doubly tagged events come from a decay chain \BdDecay, where the flavor of the \Dz can be determined using both the charge of the muon from the semileptonic $B$ meson decay, as well as from the charge of the pion from the \Dstarp meson decay. The decay chain $B^-\to \Dz \mu^- X$ is henceforth referred to as single-tag. Once the wrong-tag probability is established, the systematic uncertainty due to unrelated $\Dz\mun$ combinations is determined through pseudoexperiments.

The \Dkp  samples can be processed with the same requirements as the signal decays, since no variables related to the daughters of the \Dz meson are used in the MVA selection described in Sec.~\ref{sec:CANselection}. A weighing procedure using a gradient boosting reweighter~\cite{Rogozhnikov:2016bdp} is implemented to match the kinematics of the \Dkp samples to that of  the \Dkspp decay. Topological variables related to the $B$ meson decay including the $B$ decay vertex $\chisq$, the transverse momenta of the $\mu$ and \Dz candidates, and the pseudorapidity of the \Dz meson are used as training variables. The procedure is applied separately for single- and doubly tagged events, as the quantities related to the $B$ decay vertex differ significantly due to the different number of charged tracks used in the reconstruction. The probability of wrongly tagging the \Dkp decays is determined  as 
\begin{equation}
	\frac{R_{\rm wrong\,tag}}{1 + R_{\rm wrong\,tag}} \quad,
	\label{eqn:mistag_kpi}
\end{equation}
where $R_{\rm wrong\,tag}$ is the ratio between yields of the wrong-sign sample and right-sign sample. Wrong and right sign refer to the matching or opposite charges of the muon and kaon for single-tag events, and of the muon and pion from the \Dstarp decay for the doubly tagged events. The yields are extracted from a fit to the \Dz invariant mass distribution. The wrong-tag probability is determined for each decay-time bin separately, but is found to be time-independent and a single value is used for the full sample. The two \KS categories however need to be treated separately, as the dedicated BDTs perform differently.

The wrong-tag probabilities determined for doubly tagged \Dkspp and ${\Dkp}$ decays show good consistency. The difference is quantified as a ratio, which deviates from unity by a few per cent. The ratio is applied as a scaling factor to the wrong-tag probability obtained from the single-tag \Dkp sample, to produce the expected wrong-tag probability in the signal channel \SLDecay: $\left(0.301 \pm 0.016\right) \%$ for the DD sample and $\left(0.125 \pm 0.010\right) \%$ for the LL sample. The fraction of unrelated $\Dz\mun$ combinations is twice the measured wrong-tag rate, since such combinations have a 50\% probability to be assigned the wrong charge. An ensemble of pseudoexperiments is generated, where for the events representing an unrelated $\Dz\mun$ combination the sign of the \Dz is flipped with 50\% probability, and the $\Dz$ decay-time resolution is smeared to account for the wrong production vertex. The smearing is applied through a Gaussian of width 0.5 $\tau_{\Dz}$. The systematic uncertainty is obtained by neglecting the generated effect in the analysis of the pseudoexperiment data.

\subsection{Overall systematic uncertainties}
A summary of all the uncertainties affecting the measurement is reported in \autoref{tab:SL-syst}. The statistical
uncertainty includes, by construction, also the contribution of the uncertainties on the strong phase inputs.
The total systematic uncertainty is the sum in quadrature of the individual components. 

\begin{table}[t]
    \centering
        \caption{Summary of the uncertainties on the measured quantities. The total systematic uncertainty is the sum in quadrature of the individual components. The uncertainties due to the strong-phase inputs are (by default) included in the statistical uncertainty. Here, to ease comparison with other sources, we also report the separate contributions due to the strong phase inputs and to the statistics of the data sample. \label{tab:SL-syst}}
    \begin{tabular}{lcccc}
        \midrule
        \midrule
        Source & \xcp [$10^{-3}$] & \ycp [$10^{-3}$] & \deltax [$10^{-3}$] & \deltay [$10^{-3}$] \\
        \midrule
        
        Reconstruction and selection     & $ 0.06 $ & $ 0.79 $ & $ 0.28 $ & $ 0.24 $ \\
        Detection asymmetry              & $ 0.06 $ & $ 0.03 $ & $ 0.01 $ & $ 0.09 $ \\
        Mass-fit model                   & $ 0.03 $ & $ 0.09 $ & $ 0.01 $ & $ 0.01 $ \\
        Unrelated $\Dz\mu$ combinations  & $ 0.24 $ & $ 0.22 $ & $ 0.01 $ & $ 0.05 $ \\
        \midrule
        Total systematic  & $ 0.26 $ & $ 0.83 $ & $ 0.28 $ & $ 0.26 $  \\
        \midrule  
        \midrule  
        Strong phase inputs            & $ 0.32 $ & $ 0.68 $  & $ 0.16 $ & $ 0.21 $ \\
        Statistical (w/o phase inputs) & $ 1.45 $ & $ 3.04 $  & $ 0.92 $ & $ 1.91 $ \\
        \midrule                                                                   
        Statistical                    & $ 1.48 $ & $ 3.12 $  & $ 0.93 $ & $ 1.92 $ \\
        \midrule
        \midrule
    \end{tabular}
\end{table}

To test the robustness of the analysis, several cross-checks are performed. The analysis is repeated in subsets of data, dividing the sample by \KS categories, data-taking periods, 
magnet polarities, and kinematics of the $B$ meson. Variations of the observables \xcp, \ycp, \deltax, and \deltay measured in various subsets of data are all compatible within statistical uncertainties.
Results from the bin-flip fit are consistent with the default results when an alternative method is implemented in the decorrelation
process~\cite{Rogozhnikov:2016bdp}. Similar compatibility is observed when the selection process is altered, e.g. a different procedure is used for the multivariate analysis. These cross-checks demonstrate the  reliability and robustness of the
analysis.

\section{RESULTS}
\label{sec:Result}
The mixing and \CP-violating parameteres are obtained through a fit of the ratios of signal yields observed in regions of the Dalitz plot symmetric 
about its bisector as a function of decay-time. 
The signal yields in each Dalitz and decay-time bin are extracted using the PDFs described in Sec.~\ref{sec:ANAstrategy}. 
The widths of the signal model are fixed from a fit to the whole sample
, while the other parameters are left free to vary to account for potential mass shifts between bins due to different resonant contributions. The effect of statistical fluctuations in the low-statistics bins of the upper part of the  Dalitz plot is minimized by using the same signal PDF as determined in the fit of the corresponding bin of the higher-yield lower part of the Dalitz plot. External constraints for strong-interaction phases in each bin are used. 

\begin{figure}[!th]
\centering
\includegraphics[height=0.665\textwidth]{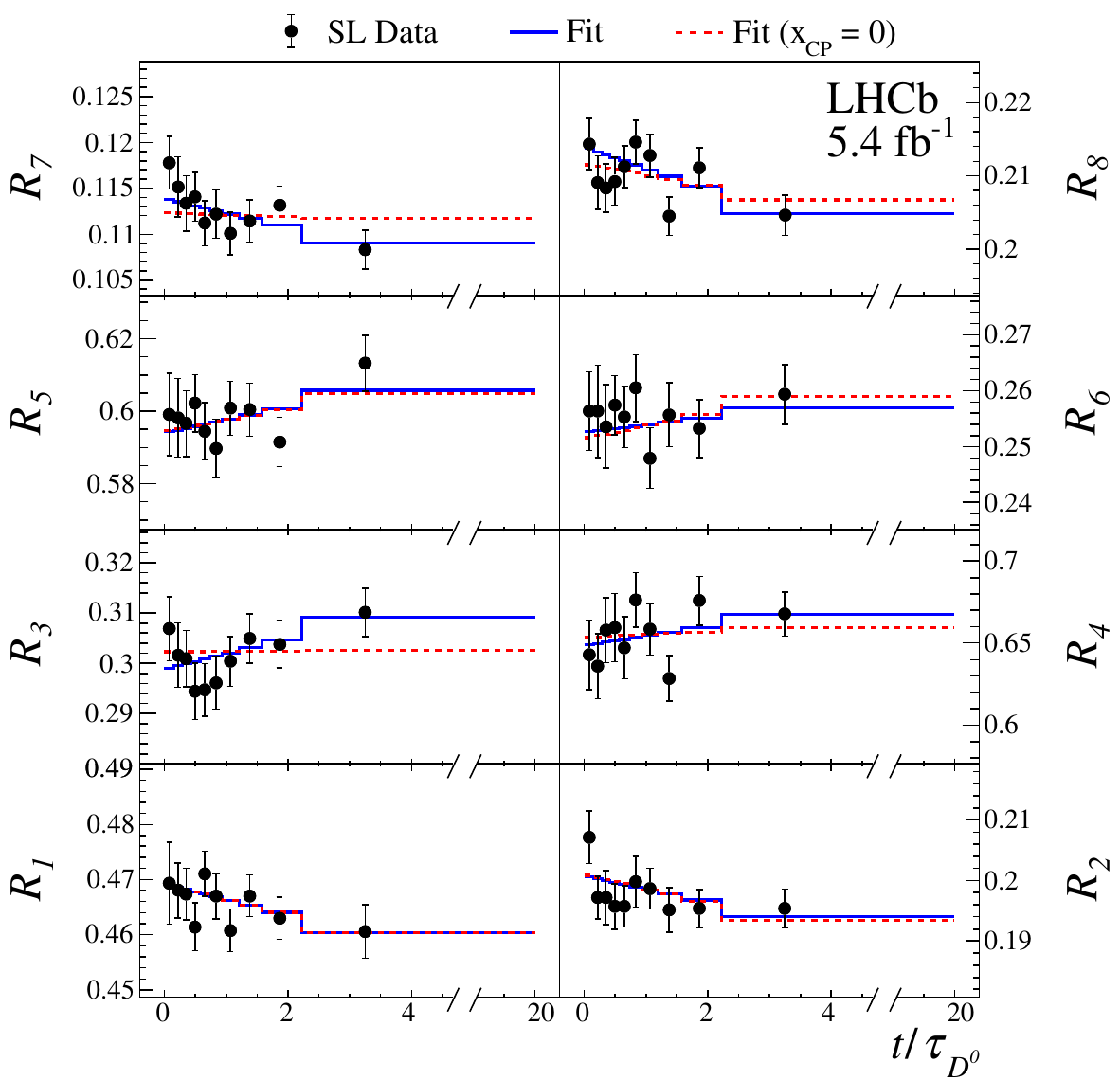}\\
\includegraphics[height=0.665\textwidth]{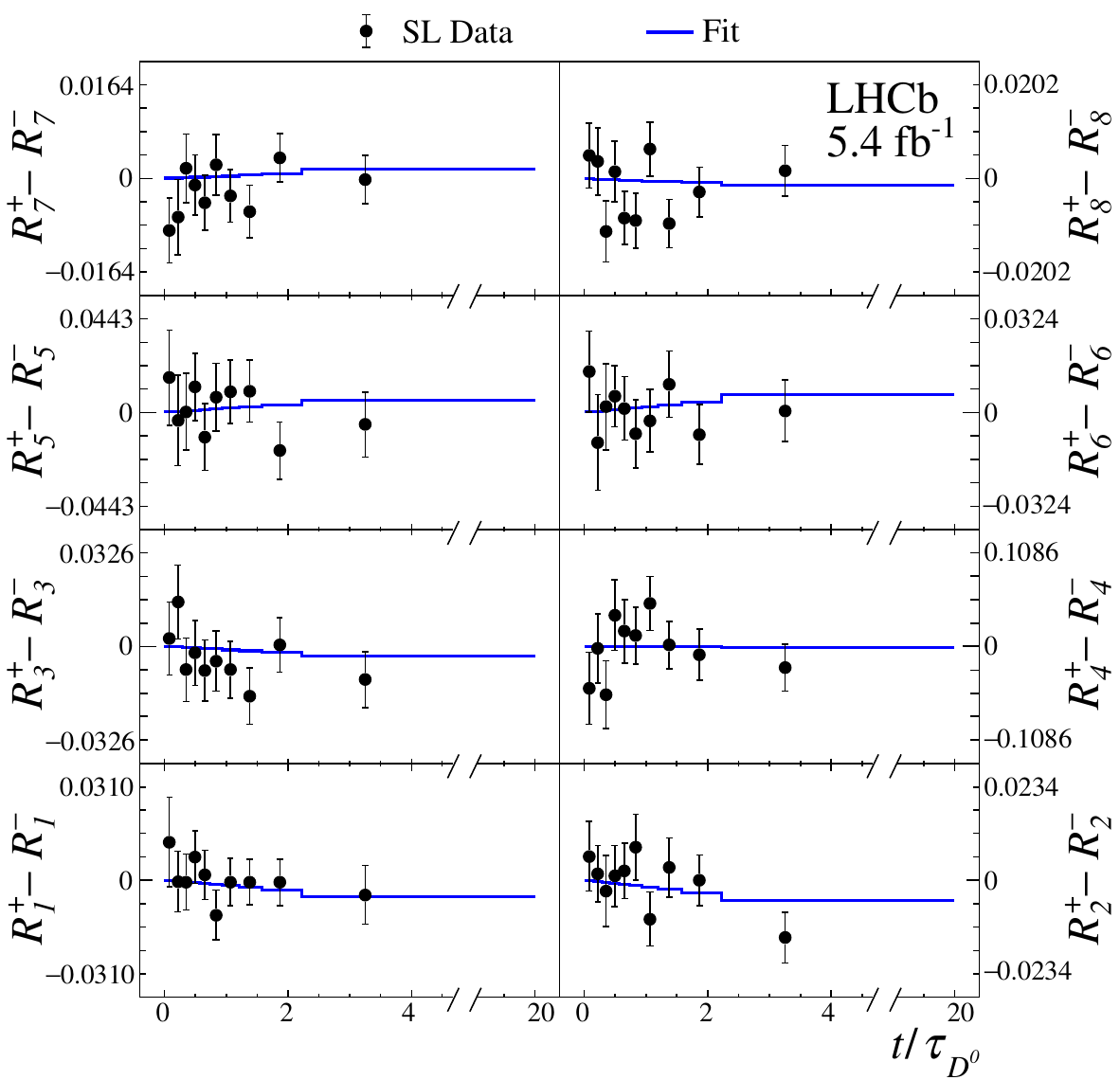}
\caption{(Top) \CP-averaged yield ratios and (bottom) difference of \Dz and \Dzb yield ratios as a function of \Dz decay-time for the different Dalitz bins. The solid blue line shows the nominal fit projections and the dashed red line shows the fit projections when \xcp is fixed to zero. \label{fig:SL-fit-projections}}
\end{figure}

The obtained results are
\begin{align*}
    \xcp    &= [  \,\,\,\,\,4.29 \pm  1.48   \pm  0.26   ]\xUnits \,, \\
    \ycp    &= [      \,\, 12.61 \pm  3.12   \pm  0.83   ]\yUnits \,, \\
    \deltax &= [  -0.77 \pm  0.93   \pm  0.28   ]\dxUnits \,,\\ 
    \deltay &= [  \,\,\,\,\,3.01 \pm  1.92   \pm  0.26   ]\dyUnits \,,
\end{align*}
where the first uncertainty is statistical and includes the contributions
due to the uncertainties of the strong phase inputs, and the second is systematic. 

Figure~\ref{fig:SL-fit-projections} shows the \CP-averaged yield ratios as well as the difference in yield ratios for \Dz and \Dzb mesons as a function of decay-time. The fit projections are shown for the nominal fit and a fit where \xcp is fixed to zero. 
The largest sensitivity to \xcp is observed, as expected, in the bins 3 and 7 where the strong phase terms, $s_3$ and $s_7$, are closest to unity.

The results are compatible with those measured in the analysis of the \PromptDecay decay~\cite{LHCb-PAPER-2021-009}. 
No \CP violation is observed.

\section{COMBINATION}
\label{sec:Combination}

As stated in Sec.~\ref{sec:ANAstrategy}, this analysis  complements the analogue analysis conducted on \PromptDecay decays, dubbed as prompt decays. 
The two analyses are statistically independent, since the overlap of events is reduced to a negligible level by selection requirements specific to each decay chain.
While the semileptonic sample has considerably fewer candidates than the prompt sample, it covers a wider \Dz decay-time: $\tau_{\rm SL}/\tau_{\Dz} \in [0, 20]$ while $\tau_{\rm prompt}/\tau_{\Dz} \in [0.3, 8]$. A combination of the two samples is therefore performed.

The systematic uncertainties from most sources can be treated as independent, with the exception of those related to detection asymmetries,
as they are estimated using the same control samples. Conservatively, a 100\% correlation is assumed for
this uncertainty. 

The combination method follows the bin-flip analysis of the data sample  from the 2011--2012 data taking campaign~\cite{LHCb-PAPER-2019-001}. 
A simultaneous minimization of a global $\chi^2$ is performed, using the prompt and semileptonic yields of subsamples separated by flavor and other 
categories (such as \KS type). The parameters $r_b$, representing the ratio of yields at $t=0$, are kept separate 
between the prompt and semileptonic samples as they are affected by different efficiencies in the Dalitz space of 
the two samples.
Allowing for \CP violation, we obtain the following averages:
\begin{align*}
    \xcp    &= [ \,\,\,\,\,4.01 \pm 0.45 \stat \pm 0.20 \syst  ]\xUnits \,, \\
    \ycp    &= [ \,\,\,\,\,5.51 \pm 1.16 \stat \pm 0.59 \syst  ]\yUnits \,, \\
    \deltax &= [ -0.29 \pm 0.18 \stat \pm 0.01 \syst  ]\dxUnits \,,\\ 
    \deltay &= [ \,\,\,\,\,0.31 \pm 0.35 \stat \pm 0.13 \syst  ]\dyUnits \,.
\end{align*}
The value of \xcp deviates from zero with a significance of $8.1\sigma$, calculated assuming Gaussian uncertainties.
There is no evidence for \CP violation.

From the results of the combination fit, \xcp, \ycp, \deltax, and \deltay are transformed into $x$, $y$, $|q/p|$, and 
$\phi$ using Eqs. (\ref{eq:xcp-def}), (\ref{eq:ycp-def}), (\ref{eq:dx-def}), and (\ref{eq:dy-def}). A likelihood 
of these parameters is constructed and confidence intervals are determined from a likelihood-ratio 
assuming that the measured correlations are independent of the true values of parameters. The
\textsc{plugin} method~\cite{plugin} is implemented for the transformation. The method is a generalization 
of the Feldman and Cousins method~\cite{Feldman:1997qc} and used in the prompt sample~\cite{LHCb-PAPER-2021-009} 
as well as the LHCb $\gamma$-combination analysis~\cite{LHCb-PAPER-2013-020}. The results are 
\begin{align*}
         x         &= (4.01\pm 0.49)\times10^{-3} \,, \\
         y         &= (\,\,\,5.5\pm 1.3\,\,\,)\times10^{-3} \,, \\
         |q/p| &= \phantom{-}1.012\,^{+\,0.050}_{-\,0.048} \,,\\ 
         \phi      &= -0.061\,^{+\,0.037}_{-\,0.044} \, \rad.
\end{align*}

\section{SUMMARY}
\label{sec:Summary}

A measurement of charm mixing and \CP-violating parameters using \Dkspp decays reconstructed in Run\,2 data, 
with the \SLDecay semileptonic decay used to identify the flavor of the charm meson at production, is presented.
The signal yields are extracted from fits to the invariant-mass distributions of the \Dz meson in bins of the Dalitz plot and \Dz decay-time. The binning of the Dalitz plot is chosen such as to preserve nearly constant values of the strong-interaction phases in each bin, 
and external constraints for these phases are used. Time-dependent ratios of yields for each pair of Dalitz plot bins symmetric about its bisector are fitted to  Eq.~(\ref{eqn:ratio}) to extract the mixing and \CP-violating parameters. The results are combined with those from the \PromptDecay analysis~\cite{LHCb-PAPER-2021-009}. Statistical and systematic correlation matrices of the measured variables are presented in the Supplemental Material~\cite{ref:SuppMat}. 

Figure~\ref{fig:Combined-comparison} shows the measured mixing and \CP-violating parameters from the ${\PromptDecay}$~\cite{LHCb-PAPER-2021-009} analysis, the \SLDecay analysis, and their combination. The combination is dominated by the result of the prompt analysis, as expected from the much larger sample size. The results obtained in this analysis are consistent with the results from the prompt analysis and the current world-average values. They represent an independent measurement and complement the knowledge of the charm mixing parameters in an extended region of the \Dz decay-time.

\begin{figure}[ht]
\centering
\includegraphics[height=0.355\textwidth]{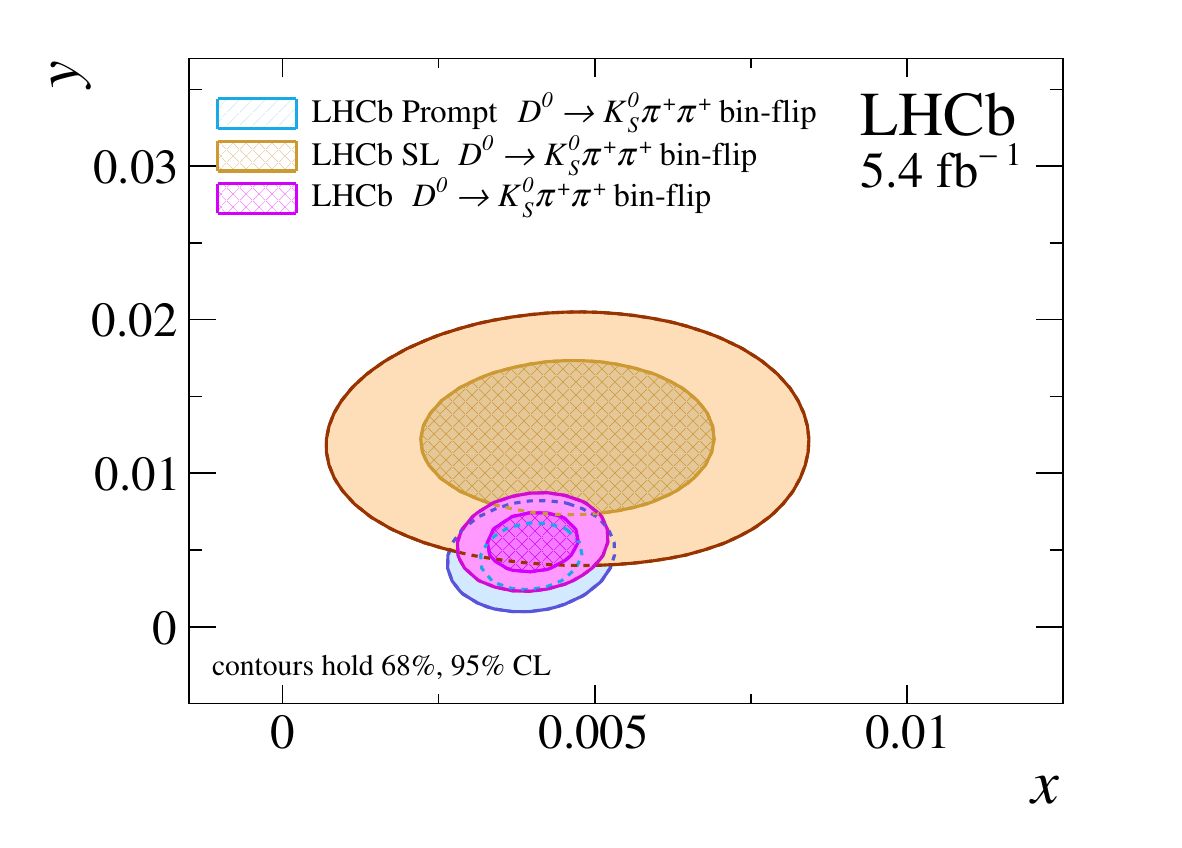}\hfil
\includegraphics[height=0.355\textwidth]{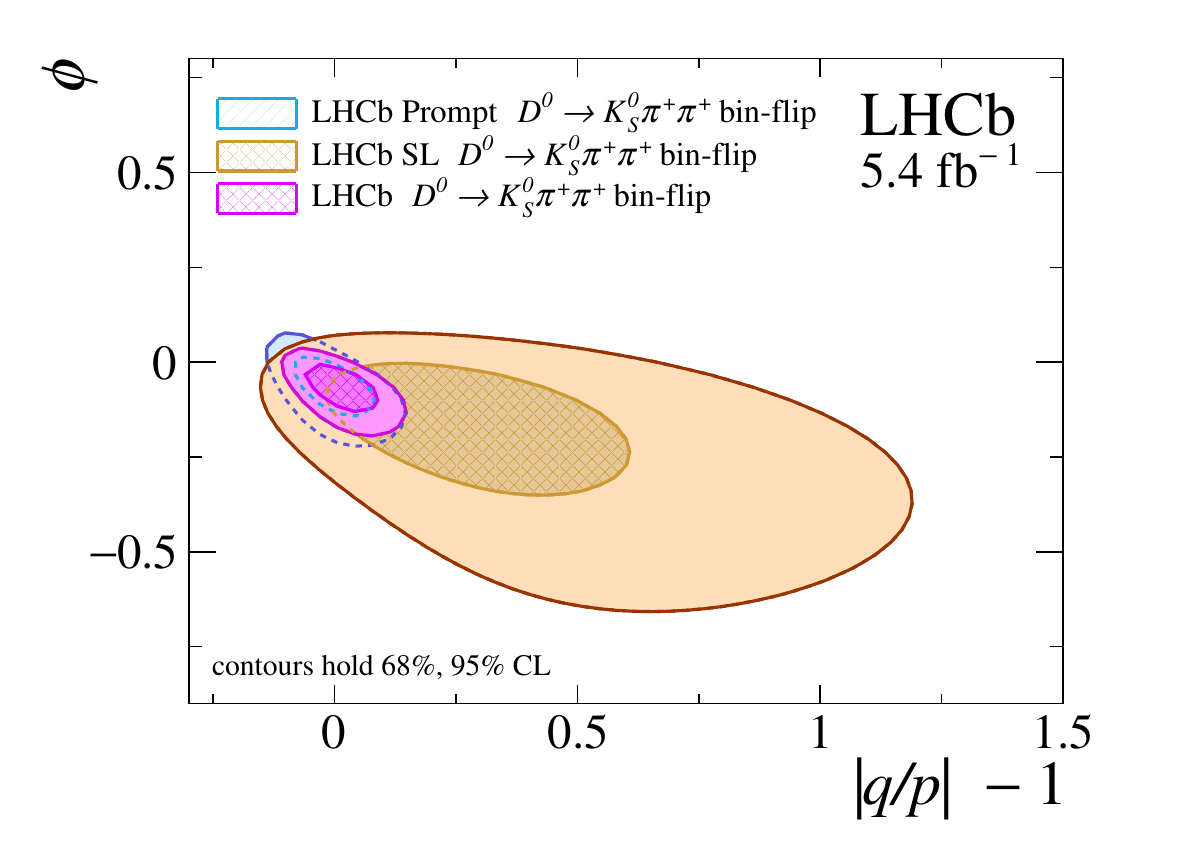}\\
\caption{Two-dimensional 68\% and 95\% confidence-level contours on (left) ($x$, $y$) and (right) ($|q/p| - 1$, $\phi$). Results from Run 2 ${\PromptDecay}$ (Prompt)~\cite{LHCb-PAPER-2021-009}, ${\SLDecay}$ (SL), and their combination are shown.   \label{fig:Combined-comparison}}
\end{figure}

\clearpage

\section*{Acknowledgements}
%
%
\noindent We express our gratitude to our colleagues in the CERN
accelerator departments for the excellent performance of the LHC. We
thank the technical and administrative staff at the LHCb
institutes.
We acknowledge support from CERN and from the national agencies:
CAPES, CNPq, FAPERJ and FINEP (Brazil); 
MOST and NSFC (China); 
CNRS/IN2P3 (France); 
BMBF, DFG and MPG (Germany); 
INFN (Italy); 
NWO (Netherlands); 
MNiSW and NCN (Poland); 
MEN/IFA (Romania); 
MICINN (Spain); 
SNSF and SER (Switzerland); 
NASU (Ukraine); 
STFC (United Kingdom); 
DOE NP and NSF (USA).
We acknowledge the computing resources that are provided by CERN, IN2P3
(France), KIT and DESY (Germany), INFN (Italy), SURF (Netherlands),
PIC (Spain), GridPP (United Kingdom), 
CSCS (Switzerland), IFIN-HH (Romania), CBPF (Brazil),
Polish WLCG  (Poland) and NERSC (USA).
We are indebted to the communities behind the multiple open-source
software packages on which we depend.
Individual groups or members have received support from
ARC and ARDC (Australia);
Minciencias (Colombia);
AvH Foundation (Germany);
EPLANET, Marie Sk\l{}odowska-Curie Actions and ERC (European Union);
A*MIDEX, ANR, IPhU and Labex P2IO, and R\'{e}gion Auvergne-Rh\^{o}ne-Alpes (France);
Key Research Program of Frontier Sciences of CAS, CAS PIFI, CAS CCEPP, 
Fundamental Research Funds for the Central Universities, 
and Sci. \& Tech. Program of Guangzhou (China);
GVA, XuntaGal, GENCAT and Prog.~Atracci\'on Talento, CM (Spain);
SRC (Sweden);
the Leverhulme Trust, the Royal Society
 and UKRI (United Kingdom).

\addcontentsline{toc}{section}{References}
\ifx\mcitethebibliography\mciteundefinedmacro
\PackageError{LHCb.bst}{mciteplus.sty has not been loaded}
{This bibstyle requires the use of the mciteplus package.}\fi
\providecommand{\href}[2]{#2}

\clearpage

\section*{Supplemental material: correlation matrices\label{supplemental-material}}
\label{sec:Supplemental}

\subsection{Correlation matrices}
Table~\ref{tab:supp-results-corrSL} presents the measured values of \SLDecay sample
together with their uncertainties and correlations. 
Table~\ref{tab:supp-results-corrComb}  shows the same information for the combined measurement. Table~\ref{tab:supp-results-corrSL-Syst}  gives correlations of each systematic effect for \SLDecay sample. Table~\ref{tab:supp-results-corrComb-Syst} shows the same information for the combined measurement. 

\begin{table}[h!]
\centering
\caption{Fit results of \xcp, \ycp, \deltax, and \deltay in \SLDecay sample. The first contribution to the uncertainty is statistical, the second systematic. Statistical and systematic correlations between \xcp, \ycp, \deltax, and \deltay are provided.}\label{tab:supp-results-corrSL}
\begin{tabular}{lcrrrrrr}
\midrule
\midrule
\multirow{2}{*}{Parameter} & Value & \multicolumn{3}{c}{Stat.\ correlations} & \multicolumn{3}{c}{Syst.\ correlations}\\
 & [$10^{-3}$] & \ycp & \deltax & \deltay & \ycp & \deltax & \deltay\\
\midrule
\xcp    & $\phantom{-}\,\,\,4.29  \pm 1.48 \pm  0.26$ & $ 0.09 $ & $ -0.01 $ & $ -0.01 $  &  $0.11$ &  $-0.25$ & $-0.02$\\
\ycp    & $\phantom{-}12.61 \pm 3.12 \pm  0.83$ &            & $  0.00 $ & $ -0.05 $  &         & $-0.05$ & $-0.20$\\
\deltax & $        \,\,\, - 0.77  \pm 0.93 \pm  0.28$ &          &           & $  0.07 $  &         &         & $0.11$\\
\deltay & $\phantom{-}\,\,\,3.01  \pm 1.92 \pm  0.26$ &          &         & \\
\midrule
\midrule
\end{tabular}
\end{table}

\begin{table}[h!]
\centering
\caption{Fit results of \xcp, \ycp, \deltax, and \deltay in the combination. The first contribution to the uncertainty is statistical, the second systematic. Statistical and systematic correlations between \xcp, \ycp, \deltax, and \deltay are provided.}\label{tab:supp-results-corrComb}
\begin{tabular}{lcrrrrrr}
\midrule
\midrule
\multirow{2}{*}{Parameter} & Value & \multicolumn{3}{c}{Stat.\ correlations} & \multicolumn{3}{c}{Syst.\ correlations}\\
 & [$10^{-3}$] & \ycp & \deltax & \deltay & \ycp & \deltax & \deltay\\
\midrule
\xcp    & $\phantom{-} 4.00 \pm 0.45 \pm 0.20$ & $ 0.12 $ & $ -0.02 $& $ -0.02 $  &  $0.08$ &  $0.00$ & $-0.01$\\
\ycp    & $\phantom{-} 5.51 \pm 1.16 \pm 0.59$ &          & $ -0.01 $& $ -0.06 $  &         & $-0.02$ & $-0.04$\\
\deltax & $           -0.29 \pm 0.18 \pm 0.01$ &          &           & $ 0.07 $  &         &         & $ 0.33$\\
\deltay & $\phantom{-} 0.31 \pm 0.35 \pm 0.13$ &          &         & \\
\midrule
\midrule
\end{tabular}
\end{table}

\begin{table}[h!]
\centering
    \caption{Correlation between \xcp, \ycp, \deltax, and \deltay for each sytematic uncertainty in the ${\SLDecay}$ sample.} \label{tab:supp-results-corrSL-Syst}
\begin{tabular}{llrrrrrr}
\midrule
\midrule
\multirow{2}{*}{Systematics} & \multirow{2}{*}{Parameter} & \multicolumn{3}{c}{Correlations} & \\
                             &                            & \ycp & \deltax & \deltay \\
\midrule
    \multirow{3}{*}{Reconstruction and selection}    & \xcp    & $ 0.18 $ & $ -0.33 $& $  0.02 $  \\
                            & \ycp    &          & $  0.00 $& $ -0.27 $  \\
                            & \deltax &          &          & $  0.10 $  \\
\midrule
    \multirow{3}{*}{Detection asymmetry}   & \xcp    & $ 0.23 $ & $ -0.48 $& $ -0.15 $  \\
                            & \ycp    &          & $ -0.10 $& $ -0.32 $  \\
                            & \deltax &          &          & $  0.17 $  \\
\midrule
    \multirow{3}{*}{Mass-fit model}    & \xcp    & $ 0.01 $ & $ -0.02 $& $  0.03 $  \\
                            & \ycp    &          & $  0.04 $& $ -0.09 $  \\
                            & \deltax &          &          & $  0.01 $  \\
\midrule
    \multirow{3}{*}{Unrelated $\Dz\mu$ combinations}    & \xcp    & $ 0.02 $ & $ -0.05 $& $  0.02 $  \\
                            & \ycp    &          & $ -0.11 $& $ -0.10 $  \\
                            & \deltax &          &          & $  0.13 $  \\
\midrule
\midrule
\end{tabular}
\end{table}

\begin{table}[h!]
\centering
\caption{Correlation between \xcp, \ycp, \deltax, and \deltay for each systematic uncertainty in the combination.}\label{tab:supp-results-corrComb-Syst}
\begin{tabular}{llrrrrrr}
\midrule
\midrule
\multirow{2}{*}{Systematics} & \multirow{2}{*}{Parameter} & \multicolumn{3}{c}{Correlations} & \\
                             &                            & \ycp & \deltax & \deltay \\
\midrule
    \multirow{3}{*}{Reconstruction and selection}    & \xcp    & $ 0.14 $ & $  0.02 $& $  0.02 $  \\
                            & \ycp    &          & $ -0.02 $& $ -0.08 $  \\
                            & \deltax &          &          & $ -0.01 $  \\
\midrule
    \multirow{3}{*}{Detection asymmetry}   & \xcp    & $ 0.33 $ & $ -0.02 $& $ -0.06 $  \\
                            & \ycp    &          & $ -0.15 $& $ -0.15 $  \\
                            & \deltax &          &          & $  0.52 $  \\
\midrule
    \multirow{3}{*}{Mass-fit model}    & \xcp    & $ 0.17 $ & $  0.00 $& $  0.01 $  \\
                            & \ycp    &          & $  0.04 $& $ -0.03 $  \\
                            & \deltax &          &          & $  0.10 $  \\
\midrule
    \multirow{3}{*}{Unrelated $\Dz\mu$ combinations}    & \xcp    & $ 0.02 $ & $ -0.05 $& $  0.02 $  \\
                            & \ycp    &          & $ -0.11 $& $ -0.10 $  \\
                            & \deltax &          &          & $  0.13 $  \\
\midrule
    \multirow{3}{*}{Secondary charm decays}    & \xcp    & $ 0.14 $ & $ -0.01 $& $  0.00 $  \\
                            & \ycp    &          & $ -0.03 $& $ -0.01 $  \\
                            & \deltax &          &          & $  0.09 $  \\
\midrule
\midrule
\end{tabular}
\end{table}

\clearpage

\subsection{Impact of external measurements of the strong phase}
\begin{table}[h!]
\centering
\caption{The statistical uncertainty on \xcp, \ycp, \deltax and \deltay, and the component of the statistical uncertainty due to the limited precision of the external
measurements of the strong phases for the semileptonic sample.}\label{tab:ext-syst}
\begin{tabular}{lcc}
\midrule
\midrule
Parameter & Statistical uncertainty [$10^{-3}$] & Component due to strong phase inputs [$10^{-3}$]\\
\midrule
\xcp    & 1.48 & 0.32\\
\ycp    & 3.12 & 0.68\\
\deltax & 0.93 & 0.16\\
\deltay & 1.92 & 0.21\\
\midrule
\midrule
\end{tabular}
\end{table}

\begin{table}[h!]
\centering
\caption{The component of the statistical uncertainty due to the limited precision of the external
measurements of the strong phases for the combined measurement of the prompt and semileptonic samples.}\label{tab:ext-syst}
\begin{tabular}{lcc}
\midrule
\midrule
Parameter & Statistical uncertainty [$10^{-3}$] & Component due to strong phase inputs [$10^{-3}$]\\
\midrule
\xcp    & 0.45 & 0.23\\
\ycp    & 1.16 & 0.66\\
\deltax & 0.18 & 0.03\\
\deltay & 0.35 & 0.05\\
\midrule
\midrule
\end{tabular}
\end{table}

\clearpage
 
\newpage
\centerline
{\large\bf LHCb collaboration}
\begin
{flushleft}
\small
R.~Aaij$^{32}$\lhcborcid{0000-0003-0533-1952},
A.S.W.~Abdelmotteleb$^{50}$\lhcborcid{0000-0001-7905-0542},
C.~Abellan~Beteta$^{44}$,
F.~Abudin{\'e}n$^{50}$\lhcborcid{0000-0002-6737-3528},
T.~Ackernley$^{54}$\lhcborcid{0000-0002-5951-3498},
B.~Adeva$^{40}$\lhcborcid{0000-0001-9756-3712},
M.~Adinolfi$^{48}$\lhcborcid{0000-0002-1326-1264},
H.~Afsharnia$^{9}$,
C.~Agapopoulou$^{13}$\lhcborcid{0000-0002-2368-0147},
C.A.~Aidala$^{77}$\lhcborcid{0000-0001-9540-4988}
% ...

{\footnotesize \it

% ...
$^{9}$Universit{\'e} Clermont Auvergne, CNRS/IN2P3, LPC, Clermont-Ferrand, France\\
% ...
$^{13}$LPNHE, Sorbonne Universit{\'e}, Paris Diderot Sorbonne Paris Cit{\'e}, CNRS/IN2P3, Paris, France\\
% ...
$^{32}$Nikhef National Institute for Subatomic Physics, Amsterdam, Netherlands\\
% ...
$^{40}$Instituto Galego de F{\'\i}sica de Altas Enerx{\'\i}as (IGFAE), Universidade de Santiago de Compostela, Santiago de Compostela, Spain\\
% ...
$^{44}$Physik-Institut, Universit{\"a}t Z{\"u}rich, Z{\"u}rich, Switzerland\\
% ...
$^{48}$H.H. Wills Physics Laboratory, University of Bristol, Bristol, United Kingdom\\
% ...
$^{50}$Department of Physics, University of Warwick, Coventry, United Kingdom\\
% ...
$^{54}$Oliver Lodge Laboratory, University of Liverpool, Liverpool, United Kingdom\\
% ...
$^{77}$University of Michigan, Ann Arbor, MI, USA, associated with Syracuse University, Syracuse, New York, USA\\
\bigskip
% ...
}
\end{flushleft}

\end{document}